\documentclass[conference]{IEEEtran}
\usepackage{cite}
\usepackage{amsmath,amssymb,amsfonts}
\usepackage{algorithmic}
\usepackage{graphicx}
\usepackage{textcomp}
\usepackage{xcolor}
\usepackage[hyphens]{url}
\usepackage{fancyhdr}
\usepackage[bookmarks=true,breaklinks=true,colorlinks,citecolor=black,linkcolor=black,urlcolor=blue]{hyperref}
\usepackage{balance}

\usepackage{dirtytalk}
\usepackage{listings}
\usepackage{multirow}
\usepackage{tikz}
\usetikzlibrary{matrix}
\usepackage{pgfplots}
\pgfplotsset{compat=1.15}
\usepgfplotslibrary{groupplots}
\usepackage{pifont}

\renewcommand{\v}[1]{\vec{#1}}

\definecolor{keywordcolor}{rgb}{0.5, 0, 0.5}
\definecolor{commentcolor}{rgb}{0.25, 0.5, 0.35}
\definecolor{stringcolor}{rgb}{0.9, 0, 0}
\definecolor{numbercolor}{rgb}{0.2, 0.2, 0.2}
\definecolor{backgroundcolor}{rgb}{0.95, 0.95, 0.92}
\title{PyPIM: Integrating Digital Processing-in-Memory from Microarchitectural Design to Python Tensors} 

\begin{document}

\author{
\IEEEauthorblockN{Orian Leitersdorf, Ronny Ronen, and Shahar Kvatinsky} \IEEEauthorblockA{\emph{Viterbi Faculty of Electrical and Computer Engineering, Technion -- Israel Institute of Technology, Haifa, Israel}} \IEEEauthorblockA{orianl@campus.technion.ac.il, ronny.ronen@technion.ac.il, shahar@ee.technion.ac.il}}

\maketitle

\pagestyle{plain}

\begin{abstract}
Digital processing-in-memory (PIM) architectures mitigate the memory wall problem by facilitating parallel bitwise operations directly within the memory. Recent works have demonstrated their algorithmic potential for accelerating data-intensive applications; however, there remains a significant gap in the programming model and microarchitectural design. This is further exacerbated by aspects unique to \emph{memristive} PIM such as partitions and operations across both directions of the memory array. To address this gap, this paper provides an end-to-end architectural integration of digital memristive PIM from a high-level Python library for tensor operations (similar to NumPy and PyTorch) to the low-level microarchitectural design.

We begin by proposing an efficient microarchitecture and instruction set architecture (ISA) that bridge the gap between the low-level control periphery and an abstraction of PIM parallelism. We subsequently propose a PIM development library that converts high-level Python to ISA instructions and a PIM driver that translates ISA instructions into PIM micro-operations. We evaluate PyPIM via a cycle-accurate simulator on a wide variety of benchmarks that both demonstrate the versatility of the Python library and the performance compared to theoretical PIM bounds. Overall, PyPIM drastically simplifies the development of PIM applications and enables the conversion of existing tensor-oriented Python programs to PIM with ease.
\end{abstract}

\section{Introduction}
\label{sec:introduction}

As the memory wall continues to limit the performance of data-intensive applications~\cite{Horowitz2014, DarkMemory}, processing-in-memory (PIM) solutions are rapidly emerging to enable logic functionality within the computer memory. The read/write memory interface is supplemented with logic operations where the CPU requests that the memory perform vectored logic on data residing at given addresses, thereby dwarfing data transfer. Whereas early proposals for PIM~\cite{ComputationalRAM, IntelligentRAM} integrated computation elements \emph{near} memory arrays, emerging PIM proposals exploit the same physical devices for both storage and logic.

% PIM figure
\begin{figure}[!t]
\centering 
\includegraphics[width=\linewidth, trim={0cm, 0.4cm, 0cm, 0cm}]{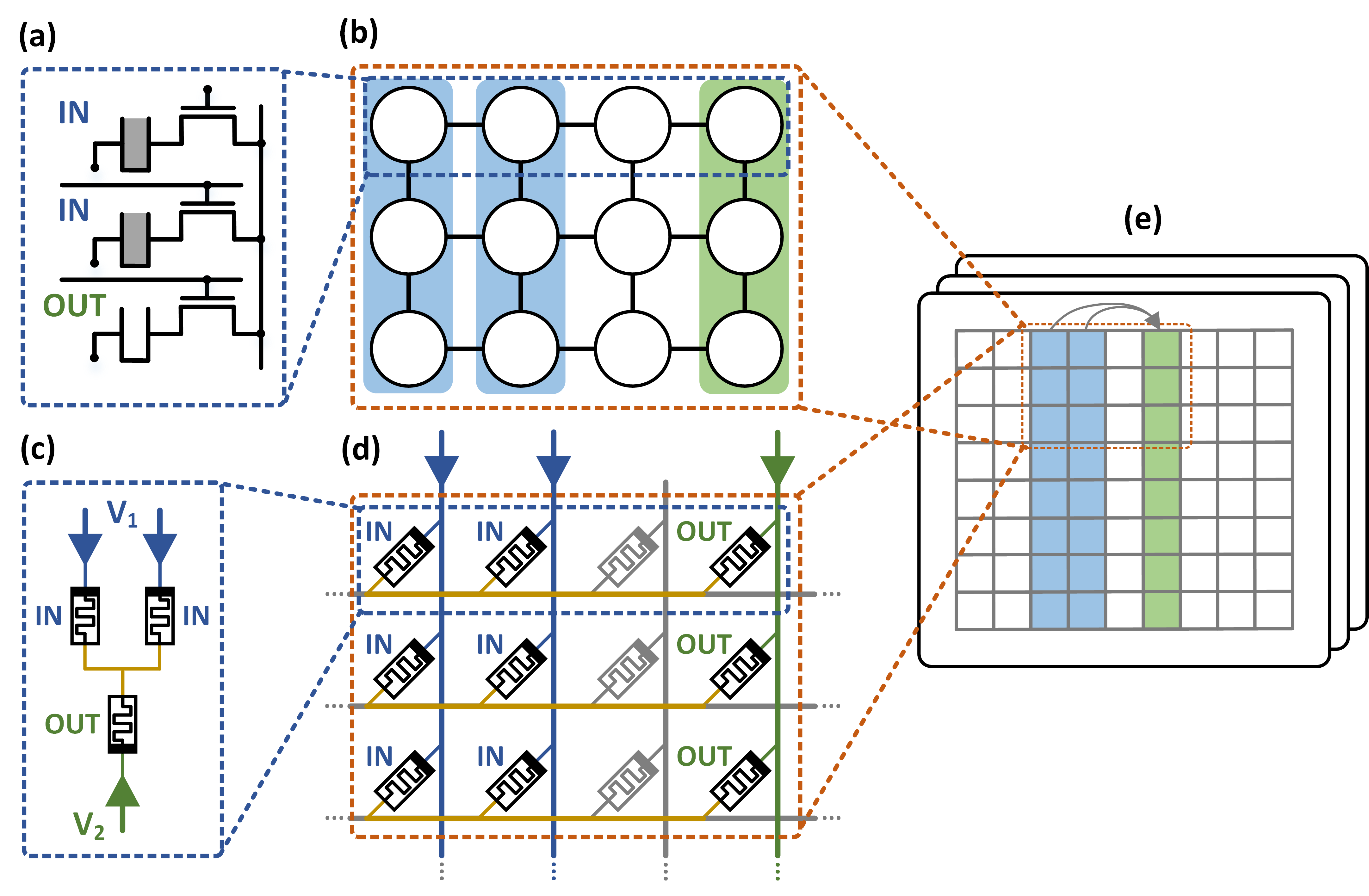}
\caption{(a) Majority logic~\cite{ComputeDRAM, SIMDRAM, DRISA, Ambit} within (b) all rows of a DRAM subarray. (c) Stateful logic~\cite{IMPLY, FELIX, MAGIC} between memristors within (d) all rows of a crossbar array. Both support (e), an abstract model enabling arbitrary bitwise operations on columns. The figure is adapted from AritPIM~\cite{AritPIM}.}
\label{fig:pim}
\end{figure}

% Overview figure
\begin{figure*}[!t]
\centering 
\includegraphics[width=\linewidth]{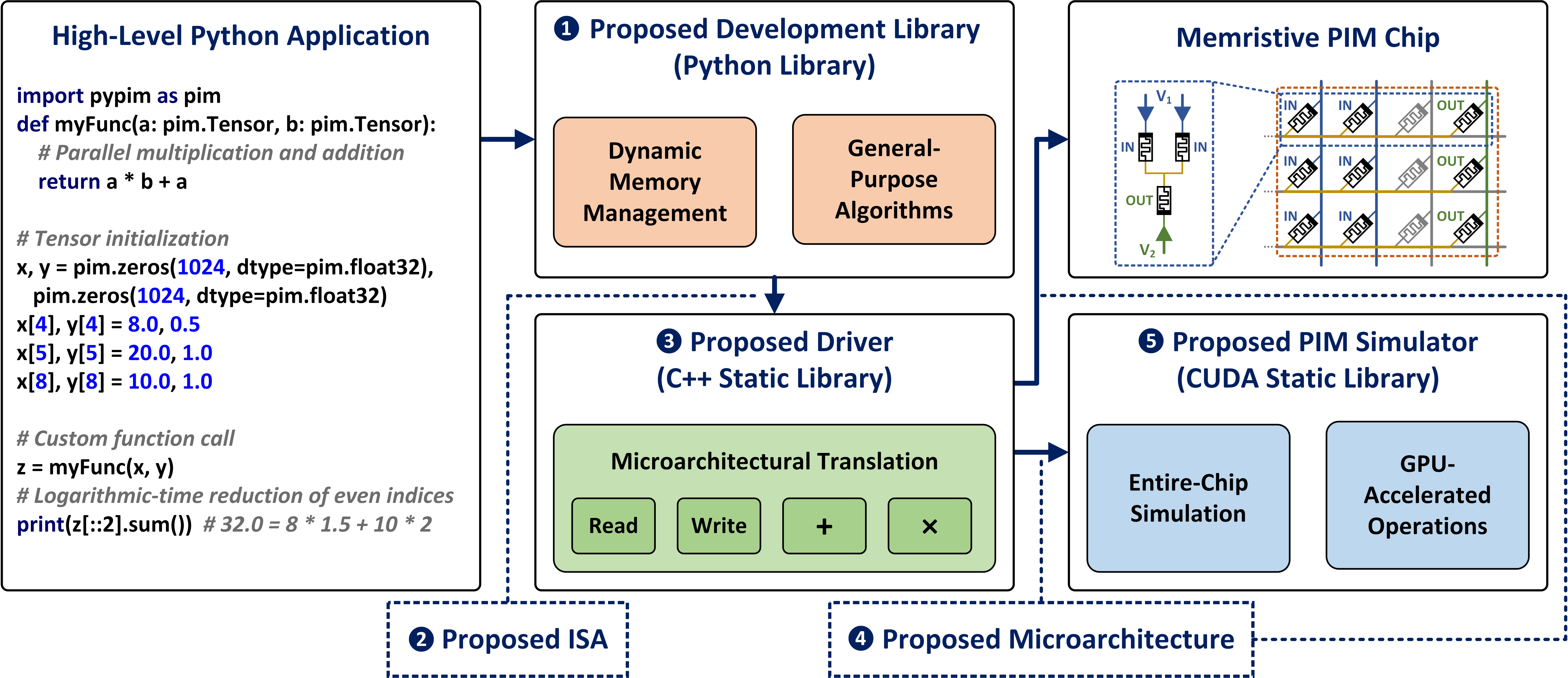}
\caption{End-to-end integration from high-level Python to the proposed microarchitecture (arrows indicate runtime dependencies), thereby enabling the development and debugging of PIM applications. The Python library utilizes syntax similar to NumPy~\cite{NumPy} for vector arithmetic (e.g., $a * b + a$), read/write operations (e.g., $x[4] = 8.0$), indexing (e.g., $z[::2]$ selects all even indices), and general-purpose routines (e.g., $.sum()$ for aggregation).}
\label{fig:overview}
\end{figure*}

Emerging digital PIM architectures (such as DRAM PIM~\cite{ComputeDRAM, SIMDRAM, DRISA, Ambit}, memristive PIM~\cite{IMPLY, FELIX, MAGIC, MemristiveLogic, Nishil, RACER} and SRAM PIM~\cite{NeuralCache, DualityCache}) enable bitwise operations within the memory with massive parallelism by designing logic circuits from the same underlying physical devices that construct the memory. For example, DRAM PIM~\cite{ComputeDRAM, SIMDRAM, DRISA, Ambit} exploits the circuit found in Figure~\ref{fig:pim}(a) to perform majority logic between the capacitors. The circuit from Figure~\ref{fig:pim}(a) is found in every column\footnote{The sub-array is illustrated transposed in Figure~\ref{fig:pim}(b).} of the DRAM subarray, thereby enabling parallel execution of the same logic gate amongst all aligned columns in the subarray and all subarrays in the memory. Similarly, memristive~\cite{Memristor} memories enable logic functionality according to the circuit in Figure~\ref{fig:pim}(c)~\cite{IMPLY, MAGIC, FELIX}, which can then similarly be repeated across all rows of a crossbar array for massive throughout~\cite{Nishil, RACER}. Unlike DRAM PIM, memristive PIM also simultaneously supports parallel operations across columns due to the symmetry of memristor arrays; further, the parallelism of memristive PIM may be increased through partitions~\cite{alamsorting, FELIX, AritPIM, MultPIM, RIME}. Therefore, in this paper, we propose a framework that supports \emph{partition-enabled memristive PIM} as the most complex case for both the low-level microarchitecture and the high-level programming model. We further consider an inter-array communication framework that consists of an H-tree hierarchy to enable distributed communication among memristor arrays.

The algorithmic challenge in the integration of digital PIM arises from the efficient exploitation of the massive bitwise parallelism. Since the gates can only be performed in parallel when they are aligned, we strive to design algorithms that maximize the gate alignment throughout the computation. The first algorithmic step involves constructing high-throughput \emph{vectored arithmetic} from the underlying basic bitwise operations. AritPIM~\cite{AritPIM} recently proposed a suite of high-throughput arithmetic operations for both fixed-point and floating-point numbers that is based on the \emph{element-parallel} approach for performing vectored arithmetic in parallel across all rows of the memory. The next step expresses applications such as matrix multiplication~\cite{MatPIM} and FFT~\cite{FourierPIM} in terms of these vectored arithmetic operations, while also manually managing the alignment of the vectors in the memory.

While significant effort has been invested in the algorithmic development of digital PIM, there have been only a few works on the underlying microarchitecture, controller design, and programming model~\cite{SIMDRAM, Nishil, RACER, DualityCache, InfinityStream}. This paper aims to provide the end-to-end architectural integration for memristive PIM that intends to bridge the gap between high-level algorithmic theory and low-level logic design. Figure~\ref{fig:overview} is an overview of this goal, providing a familiar Python development environment for tensor operations that is automatically translated into the parallel low-level operations that adhere to the proposed microarchitecture. We further propose a high-performance GPU-accelerated digital PIM simulator that models the proposed microarchitecture as a drop-in replacement for the physical chip, thereby enabling a ready-to-use platform for developing and debugging of PIM algorithms. Overall, PyPIM is the culmination of the following five components:
\begin{itemize}
    \item \textbf{Development Library (Section~\ref{sec:libraries:development}) \ding{182}:} We propose a PIM development library that enables high-level Python code (seen on the left of Figure~\ref{fig:overview}) to automatically exploit PIM parallelism. The library includes PIM-optimized dynamic memory management, intra-array and inter-array PIM algorithms, and Python operator overloading techniques that enable PIM development with ease. Furthermore, the library enables the design of new PIM routines (e.g., \verb|myFunc| in Figure~\ref{fig:overview}) using traditional Python semantics, and provides general-purpose routines such as vector reduction (summation) in logarithmic time~\cite{Bitlet} using \verb|.sum()| and efficient distributed communication for tensor alignment and shift operations.
    \item \textbf{Instruction Set Architecture (Section~\ref{sec:isa}) \ding{183}:} We propose a theoretical model for digital memristive PIM that is based on a model of warps and threads (crossbars and rows, respectively). The goal is to abstract the implementation details (e.g., the supported logic gates) while maintaining the massive throughput and flexibility of PIM. This both simplifies algorithmic development and improves generalization across PIM architectures. 
    \item \textbf{Host Driver (Section~\ref{sec:libraries:driver}) \ding{184}:} We propose an efficient driver that translates macro-instructions (instruction-set architecture) into micro-operations (micro-architecture). The driver includes elementary arithmetic routines (addition, subtraction, multiplication, and division) on fixed-point and floating-point numbers adapted from AritPIM~\cite{AritPIM}, and several new miscellaneous routines such as comparison and multiplexing to complement the suite of supported instructions. While previous works assumed that this translation is performed via a dedicated hardware controller, we designed an efficient host program that supports the PIM parallelism. This provides greater flexibility than a hardware controller as the driver may be modified in the future for additional functionality. 
    \item \textbf{Microarchitecture (Section~\ref{sec:micro}) \ding{185}:} We propose a microarchitecture for digital memristive PIM that expands the traditional read/write interface to support efficient operation decoding for partitions, flexible crossbar addressing, flexible row isolation, and hierarchical H-tree inter-crossbar communication. This is accomplished first through a variety of novel techniques that express the wide range of possible partition operations with a minimal number of bits, and then through the standardization of the operation format for the micro-operations.
    \item \textbf{GPU-Accelerated Simulator (Section~\ref{sec:evaluation}) \ding{186}:} We develop a bit-level digital PIM simulator that interfaces with the host driver as a replacement for the physical digital PIM chip. The simulator is itself GPU accelerated to reduce simulation time and enable the simulation of large-scale applications. Overall, this enables the execution, debugging, and profiling of PIM applications with ease.
\end{itemize}

This paper is organized as follows. We begin in Section~\ref{sec:background} with general background on memristive digital PIM and a summary of the related architectural works. We then continue in Section~\ref{sec:micro} with the proposed microarchitecture and in Section~\ref{sec:isa} with the proposed instruction set architecture (ISA). In Section~\ref{sec:libraries}, we propose the development and the driver libraries that enable digital PIM algorithm development with significant ease, and in Section~\ref{sec:evaluation} we evaluate the libraries using the proposed GPU-accelerated simulator on a wide variety of benchmarks. Section~\ref{sec:misc} discusses miscellaneous considerations, and Section~\ref{sec:conclusion} concludes this paper.

% ---- Background ---- %
\section{Background}
\label{sec:background}

This section begins by providing background on digital memristive PIM architectures from both the circuit and the theoretical algorithmic perspectives, and then continues with a review of previous works on the circuit and architecture considerations from various PIM types.

% ---- Digital Memristive PIM ---- %
\subsection{Digital Memristive PIM}
\label{sec:background:memristive}

Memristive PIM architectures~\cite{FELIX, Nishil, RACER, Bitlet, CRAM, FloatPIM, GraphLayout, mMPU} utilize the emerging nonvolatile memristor device~\cite{Memristor, HPMemristor} towards a dense computer memory that is inherently capable of both information storage and logic. While memristors may store multiple bits and serve as efficient analog matrix multiplication accelerators, in this work we use memristors for digital computation to achieve improved accuracy and generality~\cite{sun2023full}. This is accomplished via the stateful logic~\cite{MemristiveLogic} technique which provides the foundation for high-throughout bitwise operations.

\begin{figure}[!t]
\centering 
\includegraphics[width=\linewidth, trim={0cm, 0.1cm, 0cm, 0cm}]{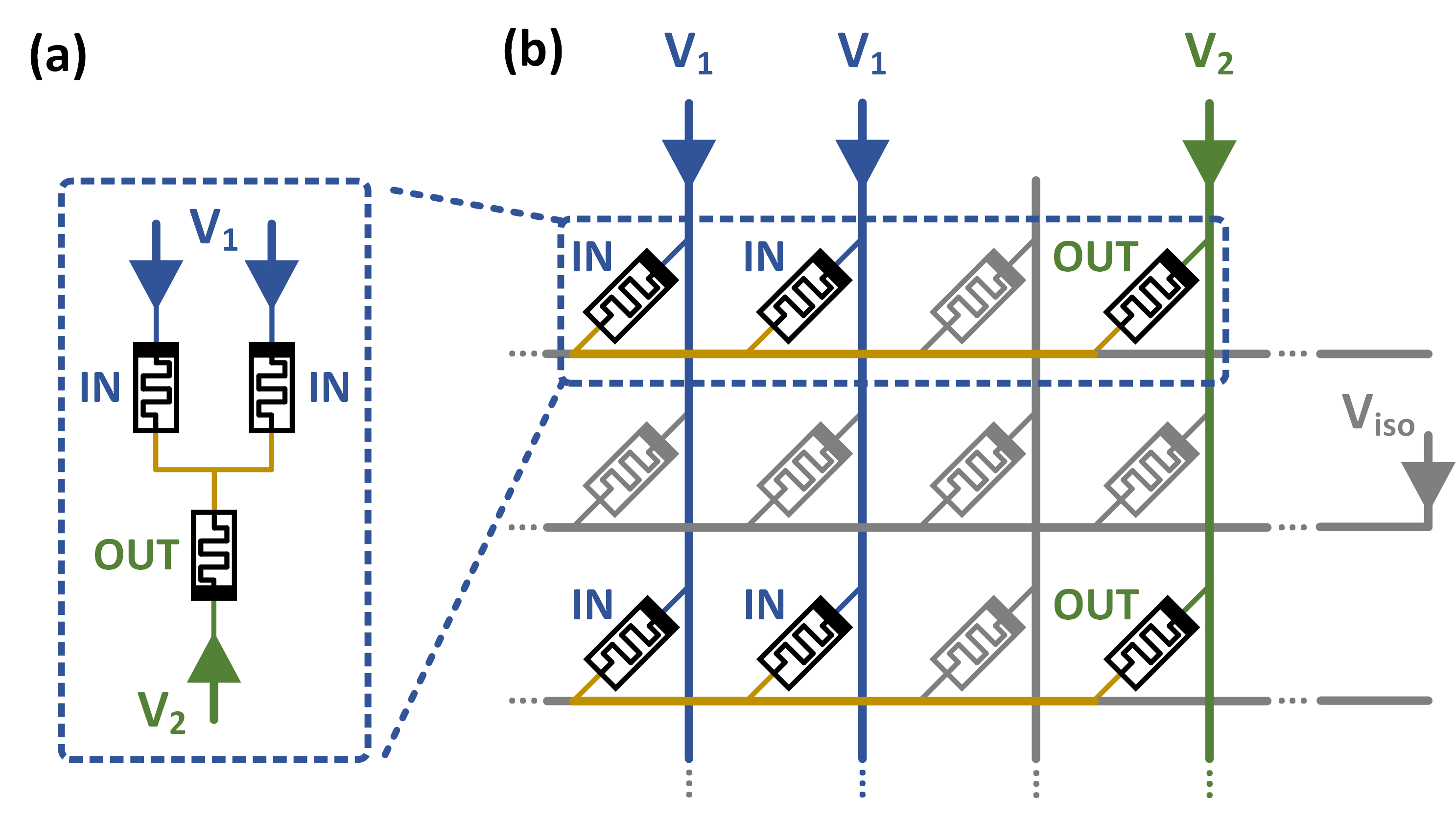}
\caption{(a) Stateful logic~\cite{MemristiveLogic} in the resistance domain between three memristors. (b) Parallel stateful logic in a crossbar array by applying ${V_1}$ and ${V_2}$ across bitlines while skipping a row, e.g., using ${V_{iso}}$.}
\label{fig:stateful}
\end{figure}

The memristor is a nonvolatile two-terminal resistive device that is inherently capable of both information storage and digital logic. Memristors possess a variable resistance that is modified through a strong current -- typically, a current in one direction may increase the resistance, whereas a current in the opposite direction decreases it. Therefore, memristors support binary information storage through their resistance, where the resistance domain is split into a binary classification (high resistance for logical zero and low resistance for logical one). Writing is performed with a relatively high current, whereas reading is performed by measuring the current from a low voltage. Furthermore, stateful logic~\cite{IMPLY, FELIX, MAGIC, MemristiveLogic} enables logic operations between memristors in the resistance domain by applying fixed voltages. Consider the circuit in Figure~\ref{fig:stateful}(a) assuming $V_1 \gg 0V$, $V_2 = 0V$, and that the output memristor is initialized to low resistance: only if the resistance of at least one of the input memristors is low (logical one), then a relatively high current will flow through the output memristor and switch it to high resistance (logical zero). That is, a logical NOR is performed between the resistance states of the input memristors, with the result being stored in the output memristor~\cite{MAGIC}. This logic technique has been extensively studied from the circuit perspective and already has several experimental demonstrations across different memristive devices~\cite{IMPLY, BarakVCM, BarakPCM, LogicComputing, StatefulLogicReview}.

% Stateful figure

% Arithmetic figure
\begin{figure*}[!t]
\centering 
\includegraphics[width=\linewidth, trim={0cm, 0.2cm, 0cm, 0cm}]{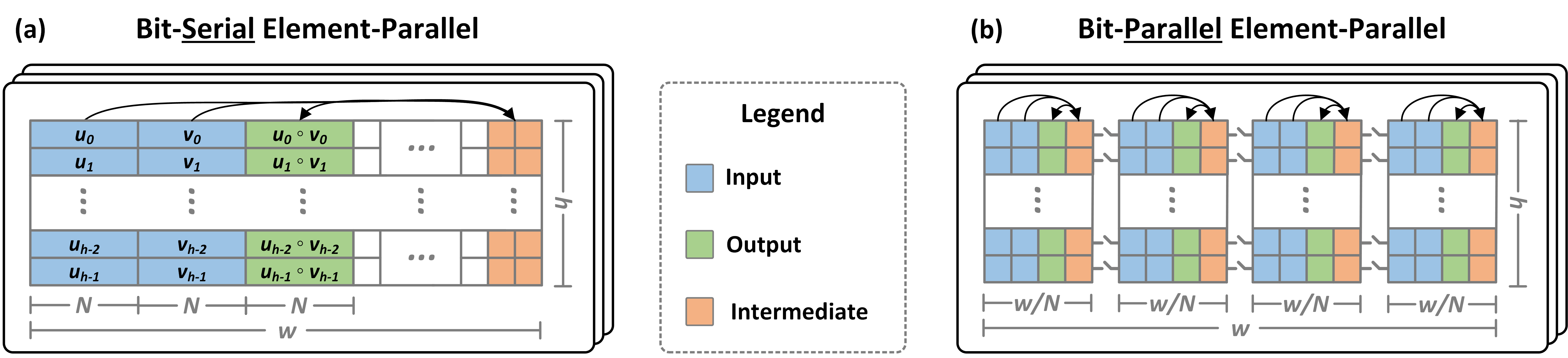}
\caption{(a) Bit-\emph{serial} element-parallel arithmetic constructs vectored arithmetic from a serial sequence of logic gates that is performed in parallel across all rows (one gate per row in every cycle). Conversely, (b) the bit-\emph{parallel} element-parallel approach stores the vectors in a \emph{strided} format across ${N}$ partitions (each bit position in a different partition) and performs up to ${N}$ gates per row per cycle. Figure adapted from AritPIM~\cite{AritPIM}.}
\label{fig:arit}
\end{figure*}

Memristive crossbar arrays enable high-density memories that also support parallel stateful logic operations. Such crossbar arrays are typically formed from a grid of, e.g., $1024 \times 1024$ memristors connected to vertical bitlines and horizontal wordlines, as seen in Figure~\ref{fig:stateful}(b). Write operations are performed by grounding a specific wordline and then applying a high voltage across the bitline(s) of the corresponding memristor(s). Read operations are performed similarly, albeit with a lower applied voltage and with a current sense amplifier connected to the bitline(s). Lastly, logic operations are performed by applying the fixed voltages $V_1$ and $V_2$ on three arbitrary bitlines, leading to the circuit seen in Figure~\ref{fig:stateful}(a) forming in all rows of the crossbar \emph{in parallel}. By broadcasting the same instruction to (up to) all of the crossbars in the overall memory, this parallelism may be extended to massive throughput for aligned bitwise operations. It is also possible to \emph{skip} (deselect) rows in stateful logic by, for example, applying an isolation voltage $V_{iso}$ to the corresponding wordline.

% ---- Element-Parallel Arithmetic ---- %
\subsection{Element-Parallel Arithmetic}
\label{sec:background:arithmetic}

The element-parallel arithmetic technique extends the bitwise parallelism towards high-throughput vectored arithmetic. Consider a crossbar of dimensions $h \times w$ containing two $h$-dimensional $N$-bit vectors $\v{u}, \v{v}$, where each row contains a single element from each vector. This technique performs a vectored operation (e.g., vector addition) on $\v{u}, \v{v}$ in parallel across all rows, in one of the following configurations:
\begin{itemize}
    \item The \emph{bit-serial} element-parallel approach~\cite{SIMPLER, Ameer, AritPIM, Nishil}, illustrated in Figure~\ref{fig:arit}(a), accomplishes this task by expressing the arithmetic operation as a serial sequence of logic gates which are executed in parallel across all rows. For example, $N$-bit addition is performed by first constructing a full-adder from 9 serial NOR gates, and then implementing ripple-carry addition in $9N$ cycles. While this approach possesses long latency, it also provides high throughout from the concurrency across rows. 
    \item Conversely, the \emph{bit-parallel} element-parallel~\cite{AritPIM, MultPIM, RIME} approach strives to provide both low latency and higher throughput by introducing dynamically-connected (transistor-based) \emph{partitions} that enable multiple gates per row per cycle. For example, with $N$ partitions, the vectors can be stored in a bit-strided format, as shown in Figure~\ref{fig:arit}(b). When the partitions are disconnected, the maximal parallelism of $N$ gates per row is possible in every cycle; the partitions may also be connected to allow gates between different partitions, enabling information transfer between them. This may accelerate arithmetic through algorithms such as carry lookahead~\cite{AritPIM} by enabling up to $N$ concurrent gates for each arithmetic operation, e.g., reducing $N$-bit multiplication latency from $O(N^2)$ to $O(N\log(N))$ ($14\times$ improvement for $N=32$~\cite{AritPIM}). Essentially, this reduces the latency from the total gate count towards the length of the critical path.
\end{itemize}

Such high-throughput arithmetic is exploited towards the acceleration of data-intensive applications. These applications can be broadly split into intra-crossbar and inter-crossbar. Intra-crossbar applications focus on extending the vector parallelism within a single crossbar to tackle more complex problems, such as matrix operations~\cite{IMAGING, FloatPIM, MatPIM} and Fast-Fourier-Transform (FFT)~\cite{FourierPIM}, where the application is primarily expressed as a sequence of vectored arithmetic operations and the data layout within the crossbar is carefully managed to guarantee data alignment. Conversely, inter-crossbar applications utilize the entire memory towards a larger application, such as neural networks~\cite{FloatPIM, ReHy}, graph operations~\cite{GraphLayout}, DNA sequencing~\cite{RAPID, FiltPIM}, and databases~\cite{PIMDB}. Inter-crossbar applications typically consist of intra-crossbar routines performed in parallel across all crossbars, supplemented with data transfer (i.e., using read and write circuitry) among crossbars via distributed communication frameworks such as H-trees. 

% ---- Related Works ---- %
\subsection{Related Work}
\label{sec:background:related}

Previous architectural research into digital memristive PIM primarily focused on designing and evaluating the peripheral circuitry rather than the programming model. Talati~et al.~\cite{NishilThesis, Nishil} and Wald~et al.~\cite{WALD201922} thoroughly investigated the circuit and peripheral considerations, including the effect of non-ideal wires on the logical correctness and the design of the peripheral circuits that apply the stateful logic voltages. Further, RACER~\cite{RACER} continued this work, evaluating additional electrical considerations and designing peripheral circuits for an entire memristive memory architecture. Therefore, in this paper, we assume the electrical and peripheral correctness from these previous works and we instead focus on the microarchitecture and its extension to the high-level tensor-based Python programming interface.

There has also been initial development of the extension to the programming interface for in-DRAM~\cite{SIMDRAM}, non-partition memristive~\cite{RACER, CONCEPT, PIMDB}, and in-SRAM~\cite{DualityCache, InfinityStream} architectures. Previously proposed ISAs~\cite{RACER, SIMDRAM, DualityCache} have abstracted bit-serial arithmetic according to a predefined set of instructions performed in parallel across several rows. Duality Cache~\cite{DualityCache} has further detailed the potential parallelism through a model similar to that of warps and threads in CUDA, and has provided a conversion from NVIDIA PTX assembly to Duality Cache instructions. Yet, these previous works do not support partition-based computation, flexible crossbar/row isolation, and computation across both directions of the array. Further, in this paper, we propose additional aspects such as (1) the dynamic memory management and mapping of aligned tensors, (2) the abstraction of inter-crossbar communication through general-purpose routines such as logarithmic reduction and tensor masking, and (3) the familiar NumPy~\cite{NumPy} syntax and algorithms provided by the Python library. Lastly, while previous works have designed on-chip controllers~\cite{SIMDRAM, RACER, PIMDB}, we propose a host driver that both provides greater flexibility (as the driver may be updated without replacing the hardware) and is not a bottleneck to PIM performance.

% ---- Microarchitecture ---- %
\section{Microarchitecture}
\label{sec:micro}

\begin{figure}[!t]
\centering 
\includegraphics[width=\linewidth, trim={0cm, 0.3cm, 0cm, 0cm}]{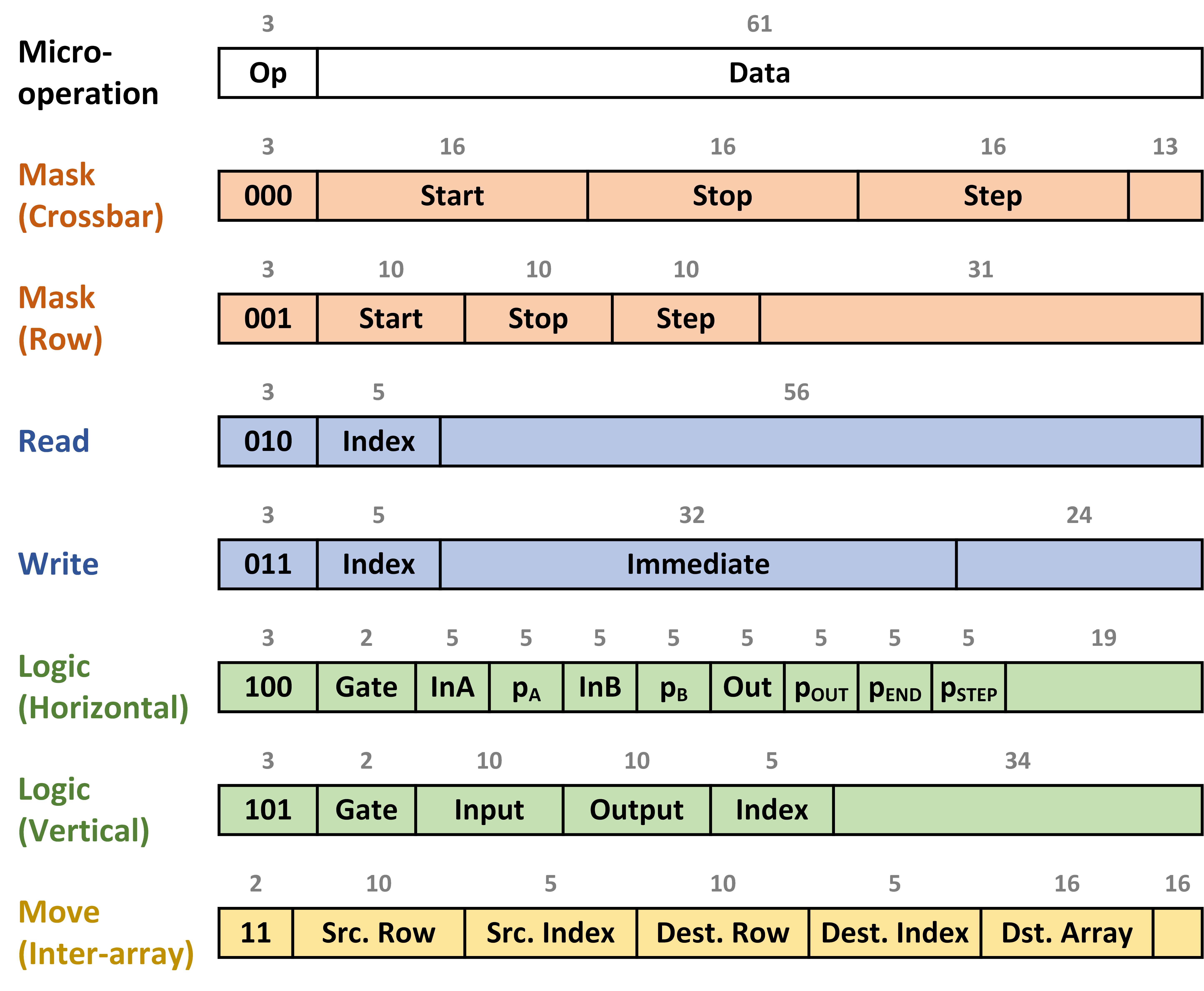}
\caption{An overview of the different proposed micro-operation types.}
\label{fig:micro-operations}
\end{figure}

This section details the proposed microarchitecture for digital memristive PIM. The microarchitecture supports efficient operation encoding for partitions, flexible range-based crossbar addressing, flexible range-based row isolation, and H-tree inter-crossbar communication. The micro-operations in this microarchitecture (referred to as \say{operations} for simplicity) are generated by the host driver proposed in Section~\ref{sec:libraries} and are broadcasted to all crossbars. As the operations directly translate into the voltages for the crossbar periphery, then the on-chip controller only needs to buffer the operations and broadcast them to the crossbars. We begin with an overview of the microarchitecture and operations in Section~\ref{sec:micro:overview} and Figure~\ref{fig:micro-operations}, and then continue by detailing each operation.

% ---- Overview ---- %
\subsection{Overview}
\label{sec:micro:overview}

% Micro-operations figure

The proposed microarchitecture supports four different operation types that enable both memory and logic functionality. For simplicity, we present here the case of $h \times w = 1024 \times 1024$ crossbar size with $N=32$ partitions comprising an 8GB memory (64k crossbars)~\cite{AritPIM, Bitlet} that supports NOT and NOR operations in the horizontal direction and NOT operations in the vertical direction. Regardless, the libraries provided in Section~\ref{sec:libraries} can be configured according to different parameters and gates, there are sufficient unused bits in the format for larger memories, and the proposed mechanisms can be generalized to the case where the number of partitions differs from $N$ (the size of a word in the architecture). The microarchitecture interface consists of 64-bit operations sent from the host driver, with an optional $N$-bit response for read operations. The supported operation types are:
\begin{enumerate}
    \item \emph{Mask:} These set the per-crossbar and per-row masks, indicating which rows are active in the next operations. 
    \item \emph{Read/Write:} Standard read/write operations to the memory with $N$-bit granularity. The target crossbar(s) and row(s) are specified in preceding \emph{mask} operations, and then the intra-row index is specified in the operation.
    \item \emph{Logic:} These operations communicate a logic operation and are split into \emph{horizontal} operations (as seen in Figure~\ref{fig:stateful}) that encode partition operations and \emph{vertical} operations that essentially transfer data between two rows (e.g., using a NOT gate in the transposed direction).
    \item \emph{Move:} These operations communicate a parallel \emph{distributed} inter-array data movement among the arrays in an H-tree hierarchical structure. 
\end{enumerate}

% ---- Mask Operations ---- %
\subsection{Mask Operations}
\label{sec:micro:mask}

These operations select the rows of the memory that will be activated in the following read/write and logic operations. While the maximal PIM throughput is attained when all crossbars and all rows participate in the computation, it may be required to only operate on selected crossbars or rows (e.g., using isolation voltages, see Section~\ref{sec:background:memristive}) to either avoid corrupting unselected data or reduce energy consumption. We support a range-based pattern for these masks, defined according to $start, stop$, and $step$ values as $\{start, start + step, start + 2 \cdot step, \hdots, stop\}$ (where they must satisfy that $step$ divides $stop - start$). This pattern enables the flexibility required by previous PIM applications works and requires a small representation size.

The crossbar mask is implemented with every crossbar storing a single volatile bit representing whether that crossbar is currently activated. Whenever a crossbar mask operation is broadcasted, the peripheral circuitry of every crossbar updates the stored activation bit accordingly. For the following non-mask operations, the stored activation bit acts as an enable bit for the entire operation (i.e., the operation is not performed in the given crossbar if the stored activation bit is false).

The row mask is implemented by the crossbars storing the $start, stop$, and $step$ values and utilizing them in non-mask operations. The row mask is updated by a row mask operation that applies to all crossbars and provides the updated $start, stop$, and $step$ values. During read/write and horizontal logic operations, the row mask is expanded into a binary vector of length $h$ representing the activated rows and is then used as the enable bits for the desired operation (for example, whether to apply $V_{iso}$ across rows for stateful logic operations).

% ---- Read/Write Operations ---- %
\subsection{Read/Write Operations}
\label{sec:micro:rw}

% Read figure
\begin{figure}[!t]
\centering 
\includegraphics[width=\linewidth, trim={0cm, 0.3cm, 0cm, 0cm}]{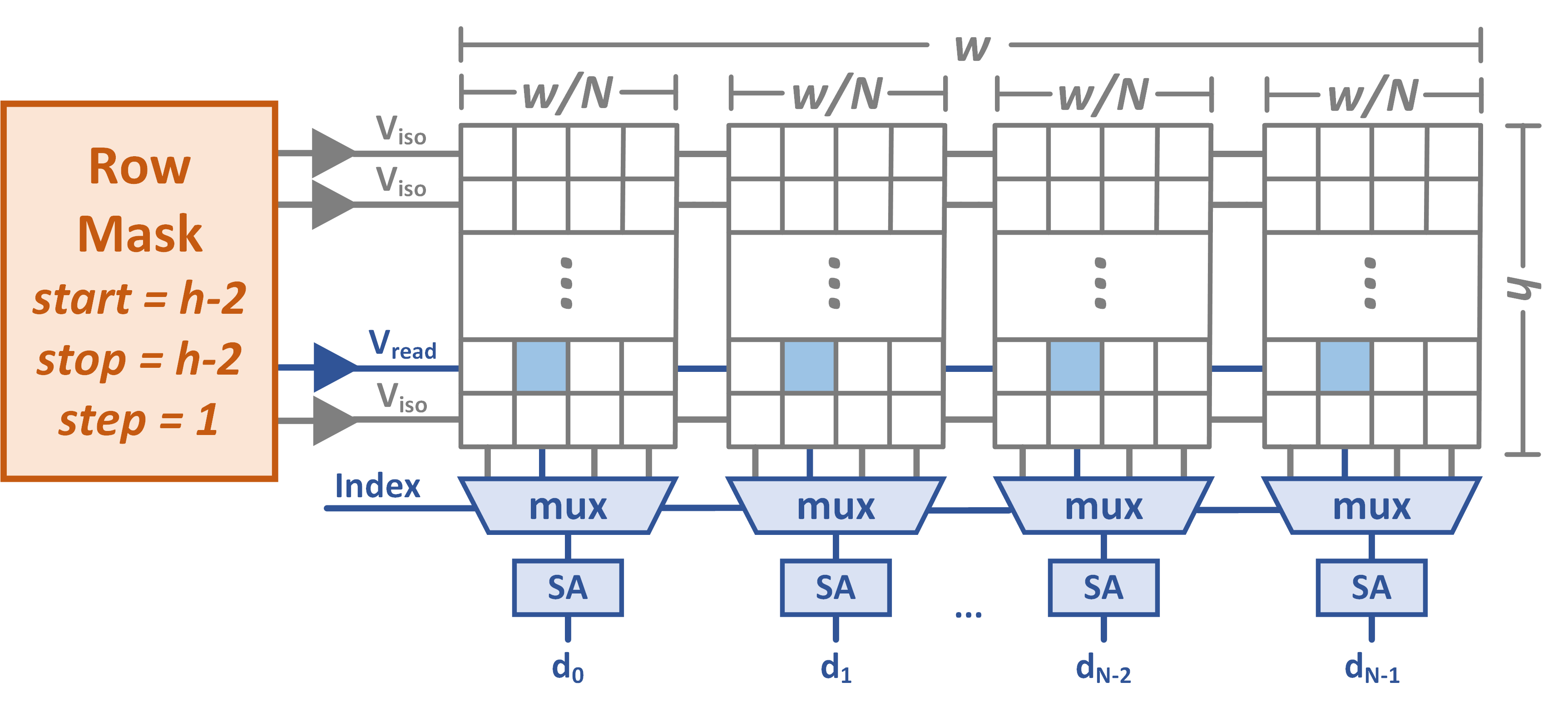}
\caption{An overview of the strided data access of the reading mechanism~\cite{NishilThesis}.}
\label{fig:read}
\end{figure}

% Parallelism figure
\begin{figure*}[!t]
\centering 
\includegraphics[width=\linewidth, trim={0cm, 0.2cm, 0cm, 0cm}]{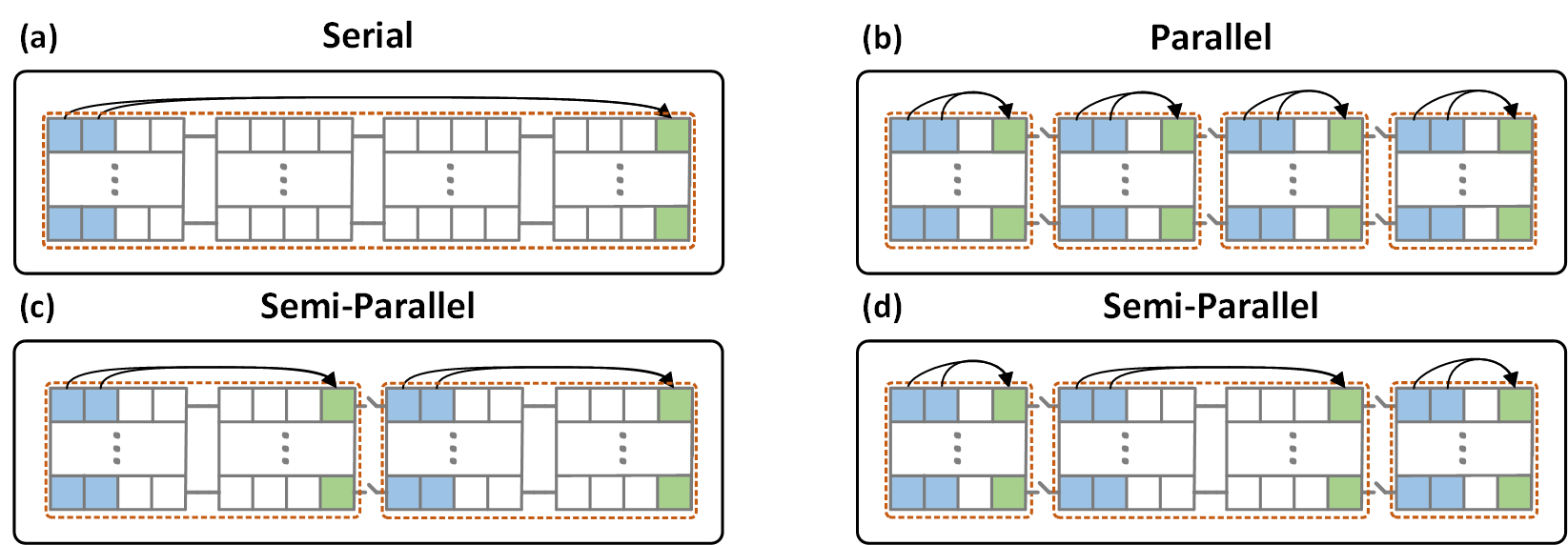}
\caption{Overview of the different types of partition-based parallelism: (a) serial, (b) parallel, and (c,d) semi-parallel. The dynamic section division is shown in dashed orange, inputs are blue, and outputs are green.}
\label{fig:parallelism} 
\end{figure*}

The read/write operations enable standard memory access in addition to the PIM functionality. The operations access the data in $N$-bit granularity in a strided memory format, where preceding mask operations first select the desired crossbar(s) and the row(s), and then the read/write operation specifies the intra-row index. Read operations are performed following mask operations that select a single row in a single crossbar and further provide a $\log(w/N)$-bit index specifying the intra-row strided address. Figure~\ref{fig:read} exemplifies this case for $N=4$ where row $h-1$ is selected at index $2$. The strided memory access is due to the sense amplifiers being shared amongst several consecutive bitlines to reduce area~\cite{NishilThesis, RACER}, yet this also coincides with the strided memory format utilized in bit-parallel element-parallel computation as there is a single multiplexer for each partition. Write operations are performed similar to Figure~\ref{fig:read}, yet the mask may select multiple crossbars and rows to enable parallel write operations for the same data.

% ---- Logic Operations ---- %
\subsection{Logic Operations (Horizontal)}
\label{sec:micro:hlogic}

We propose an operation format for encoding partition operations. We begin in Section~\ref{sec:micro:hlogic:parallelism} with a review of the different forms of partition parallelism (serial, semi-parallel, and parallel) from the perspective of the computational model~\cite{alamsorting, AritPIM, MultPIM, RIME}, continue in Section~\ref{sec:micro:hlogic:half} by proposing the \emph{half-gates} technique which supports this parallelism with standard crossbar periphery, and then conclude in Section~\ref{sec:micro:hlogic:minimal} with the overall proposed operation format that simultaneously provides the flexibility required by previous algorithmic works and a relatively-small encoding size.

% ---- Partition Parallelism ---- %
\subsubsection{Partition Parallelism}
\label{sec:micro:hlogic:parallelism}

Partitions enable a unique parallelism that may be exploited for efficient arithmetic techniques. Consider inserting $N-1$ transistors at fixed locations into each row of an $h \times w$ crossbar, as illustrated in Figure~\ref{fig:parallelism}. The transistors dynamically isolate different parts of each row to enable concurrent execution, essentially dynamically dividing the crossbar partitions into \emph{sections} (dashed orange) such that each section may perform a parallel stateful logic operation. Initial works~\cite{FELIX, RIME, alamsorting} utilized partitions in a binary fashion: either the entire crossbar is one section (serial -- see Figure~\ref{fig:parallelism}(a)) or each partition is a section (parallel -- see Figure~\ref{fig:parallelism}(b)). Recent works demonstrated the potential of semi-parallelism, e.g., a further $4\times$ improvement for multiplication~\cite{AritPIM, MultPIM} compared to the previous binary solution~\cite{RIME}. We define these parallelism forms, presenting a trade-off between parallelism (gates per cycle per row) and flexibility (inputs and outputs from different partitions to facilitate communication between bit positions):
\begin{itemize}
    \item \emph{Serial} (Figure~\ref{fig:parallelism}(a)): When the transistors are \emph{all conducting}, the crossbar is equivalent to one without partitions. Thus, only a single gate is executed per row per cycle. 
    \item \emph{Parallel} (Figure~\ref{fig:parallelism}(b)): When the transistors are \emph{all not conducting}, then $N$ gates may operate concurrently as part of an operation, \emph{one gate within each partition}, or $N$ gates per row per cycle. 
    \item \emph{Semi-Parallel} (Figures~\ref{fig:parallelism}(c,d)): When only \emph{some} transistors are \emph{conducting}, then multiple gates may operate concurrently, \emph{between partitions}. Essentially, the sections that define the concurrent gates must not intersect.
\end{itemize}

% ---- Half-Gates Technique ---- %
\subsubsection{Half-Gates Technique}
\label{sec:micro:hlogic:half}

\begin{figure*}[!th]
\centering 
\includegraphics[width=\linewidth, trim={0cm, 0.2cm, 0cm, 0cm}]{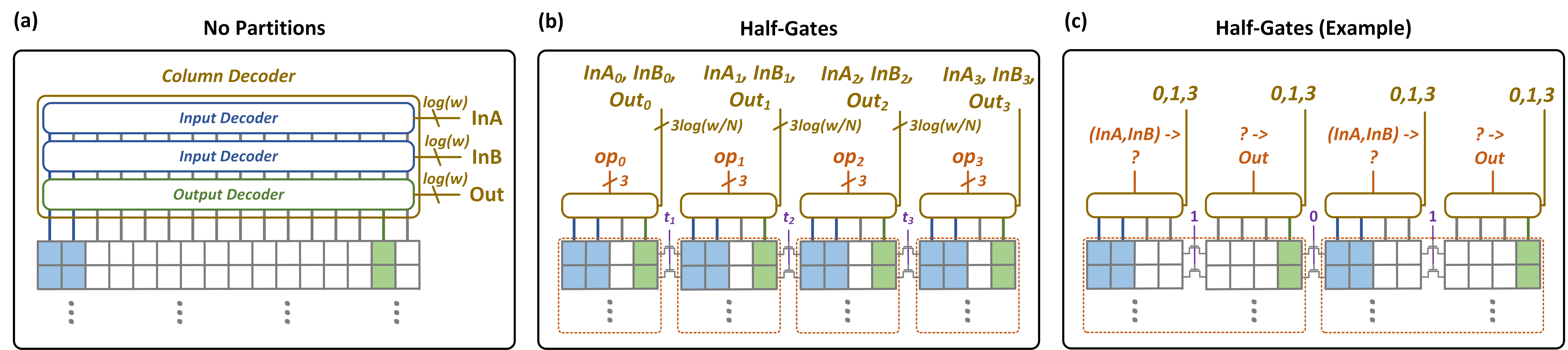}
\caption{(a) The stateful logic periphery for a baseline crossbar with no partitions~\cite{NishilThesis}. (b) The periphery as proposed by the \emph{half-gates} technique, with (c) an example for semi-parallelism corresponding to Figure~\ref{fig:parallelism}(c).}
\label{fig:flexible}
\end{figure*}

This section provides the core technique utilized in PyPIM to support the abstract parallelism provided by partitions using standard crossbar periphery components. The traditional operation format for horizontal logic in non-partition crossbars includes three-column indices that specify the input and output bitlines. Further, the operation also specifies the gate type to be performed; similar to previous works~\cite{AritPIM, Nishil, RACER}, without loss of generality, we assume four operations of $\{INIT0, INIT1, NOT, NOR\}$, where $INITx$ is a constant logic gate without inputs that sets the output to $x$ (similar to a write operation). The periphery for such a crossbar consists of a column decoder that receives the indices $InA$, $InB$, and $Out$, and applies $V_{1}$ on bitlines $InA$/$InB$ and $V_{2}$ on bitline $Out$, see Figure~\ref{fig:flexible}(a); the voltages $V_1$ and $V_2$ are chosen according to the gate. The column decoder consists of three decoders, each of which receives a column index and outputs either $V_1$ or $V_2$~\cite{mMPUController, MemristiveLogic, NishilThesis}.

\begin{table}[!t]
\centering
\caption{Per-partition opcodes in the half-gate technique.}
\begin{tabular}{c|c|c|c}
 \textbf{Index} & \textbf{Opcode} & \textbf{Index} & \textbf{Opcode} \\
 \hline \hline
 000 & - & 100 & $(InA, ?) \rightarrow \; ?$ \\
 001 & $? \rightarrow Out$ & 101 & $(InA, ?) \rightarrow Out$ \\
 010 & $(?, InB) \rightarrow \; ?$ & 110 & $(InA, InB) \rightarrow \; ?$ \\
 011 & $(?, InB) \rightarrow Out$ & 111 & $(InA, InB) \rightarrow \; Out$ \\
\end{tabular}
\label{table:opcodes}
\end{table}

Our proposed approach is based on a novel technique of \emph{half-gates}: we utilize a single-column decoder per partition, and introduce per-partition opcodes that are exploited towards partition parallelism, as shown in Figure~\ref{fig:flexible}(b). We describe the basic idea through the following example: to support a stateful logic gate where inputs are in partition $p_a$ and outputs are in partition $p_b$ (both in the same section), (1) the column decoder of $p_a$ applies only the input voltages without applying the output voltages, and (2) the column decoder of $p_b$ applies only the output voltages without applying the input voltages. Essentially, $p_a$ \say{trusts} that a different partition will apply output voltages, and $p_b$ \say{trusts} that a different partition will apply input voltages. While each gate on its own is not valid (\emph{half-gate}), their combination is valid. Table~\ref{table:opcodes} details the various possible opcodes for each column decoder, where \say{?} represents not applying voltages for that part of the gate and \say{-} represents not applying voltages at all (e.g., for partitions in between $p_a$ and $p_b$). An example is shown in Figure~\ref{fig:flexible}(c) for the operation from Figure~\ref{fig:parallelism}(c). The opcode decoding utilizes the first two bits as the enable bits for the input decoders, and the last bit as the enable bit for the output decoder.

% ---- Partition Model ---- %
\subsubsection{Partition Model}
\label{sec:micro:hlogic:minimal}

While the model proposed in the previous section (and highlighted in Figure~\ref{fig:flexible}(b)) can support the full flexibility of semi-parallelism from Section~\ref{sec:micro:hlogic:parallelism}, this would lead to a massive operation encoding due to the vast number of partition operations possible. Therefore, this section proposes three restrictions on the model that enable a compact operation format that still supports the previous algorithmic PIM works as they already adhere to these patterns. 

The first restriction requires identical \emph{intra-partition} indices; that is, $InA_0 = InA_1 = \cdots, InB_0 = InB_1 = \cdots, Out_0 = Out_1 = \cdots$. The example in Figure~\ref{fig:flexible}(c) already satisfies this requirement as the same $InA,InB,Out = 0,1,3$ are inputted to all of the column decoders. Notice that, as in the case of Figure~\ref{fig:flexible}(c), some of the operands are not utilized by the partitions (e.g., a partition with opcode \say{$(InA, InB) \rightarrow\; ?$} will not use $Out$), yet this does not affect the correctness.

The second restriction is that the opcodes repeat periodically. Consider the opcodes representing the leftmost gate (e.g., the opcodes of \say{$(InA,InB) \rightarrow\; ?$} and \say{$? \rightarrow Out$} in the first and second partitions of Figure~\ref{fig:flexible}(c), respectively). We require that the opcodes for the remaining concurrent gates in the operation repeat; e.g., the opcodes in the third and fourth partitions in Figure~\ref{fig:flexible} are a repetition of the opcodes in the first and second partitions. The operation encoding includes (1) the partition indices $p_A, p_B, p_{OUT}$ of the two inputs and the output of the leftmost gate (where $p_A \leq p_B$), and (2) the periodicity represented by the index of the partition containing the output of the last gate $p_{END}$ and the step size $p_{STEP}$.

The third restriction aims to deduce the transistor selects from the opcodes. The partition model as defined above enabled some flexibility in the selection of the transistor selects for the same set of gates. For example, if there are two concurrent gates, one from partition $0$ to partition $3$ and another from partition $8$ to partition $11$, then any one of the transistors between partition $3$ and partition $8$ can be set to non-conducting for the operation to be valid. For the case of $p_A \leq p_{OUT}$ ($p_A > p_{OUT}$ is similar), we restrict the transistor selects to adhere to the following pattern: a transistor is set to non-conducting only if the partition to its left has opcode $* \rightarrow Out$ (where $*$ reflects \say{don't care}) or the partition to its right has opcode $(InA, *) \rightarrow *$. This restriction enables the generation of the transistor selects from other existing fields.

Overall, we find that this operation format requires $2 + 3\cdot \log(w) + 2 \cdot \log(N) = 42$ bits in total (gate type, input/output indices, and opcode pattern), only a $1.31\times$ increase over that of a crossbar without partitions. Figure~\ref{fig:micro-operations} details this format, providing $19$ unused bits out of the available 64 bits. We find that this restricted partition model still supports the full flexibility of the previous algorithmic works as it supports the underlying fundamental routines utilized in those works. Previous works that have utilized the high flexibility of semi-parallelism have first designed useful general-purpose partition techniques such as broadcast and reduction operations~\cite{MultPIM, AritPIM}, and have then utilized those routines to tackle larger problems such as parallel prefix carry-lookahead addition~\cite{AritPIM}. As the operations utilized in the general-purpose partition techniques adhere to the minimal partition model, we find that this model may also support the more complex arithmetic algorithms. We demonstrate this in Section~\ref{sec:libraries} by implementing the AritPIM~\cite{AritPIM} suite using the proposed microarchitecture. 

% ---- Logic Operations ---- %
\subsection{Logic Operations (Vertical)}
\label{sec:micro:vlogic}

Stateful logic also supports operations in the transpose direction when the voltages $V_1, V_2$ are applied on the wordlines rather than the bitlines~\cite{Nishil}. These operations are primarily utilized to transfer data \emph{between different rows of the same array}, such as using two consecutive NOT gates~\cite{Bitlet}, as arithmetic in the transpose direction is not possible \emph{when $N$-bit numbers are stored across $N$ horizontal cells}. Therefore, we support only the $\{INIT0, INIT1, NOT\}$ set of gates for vertical stateful logic operations. The operation format includes the input and output rows for the vertical gate, as well as the $Index$ field that represents the column mask. That is, the vertical logic operation is applied to the columns that are at an intra-partition index equal to $Index$, similar to the access pattern of read/write operations. 

% ---- Move Operations ---- %
\subsection{Move Operations (Inter-Array)}
\label{sec:micro:move}

We propose an inter-array communication framework based on a hierarchical H-tree structure connecting the crossbars. Previous works~\cite{RACER, FloatPIM, PIMDB} have required data movement within and between crossbar arrays for the acceleration of data-intensive applications. Notice that intra-crossbar data transfer (see Section~\ref{sec:micro:vlogic}) is supported with massive parallelism since it may occur in parallel across all crossbar arrays, yet achieving inter-crossbar data transfer with the above operations (i.e., perform a read operation from one crossbar and then a write operation to the other crossbar) leads to entirely serial data movement that reaches the host processor. Therefore, we seek to provide intermediate levels of data transfer that enable some distributed inter-crossbar communication with parallelism across pairs of crossbars. 

\begin{figure}[!t]
\centering 
\includegraphics[width=0.85\linewidth, trim={0cm, 0.2cm, 0cm, 0cm}]{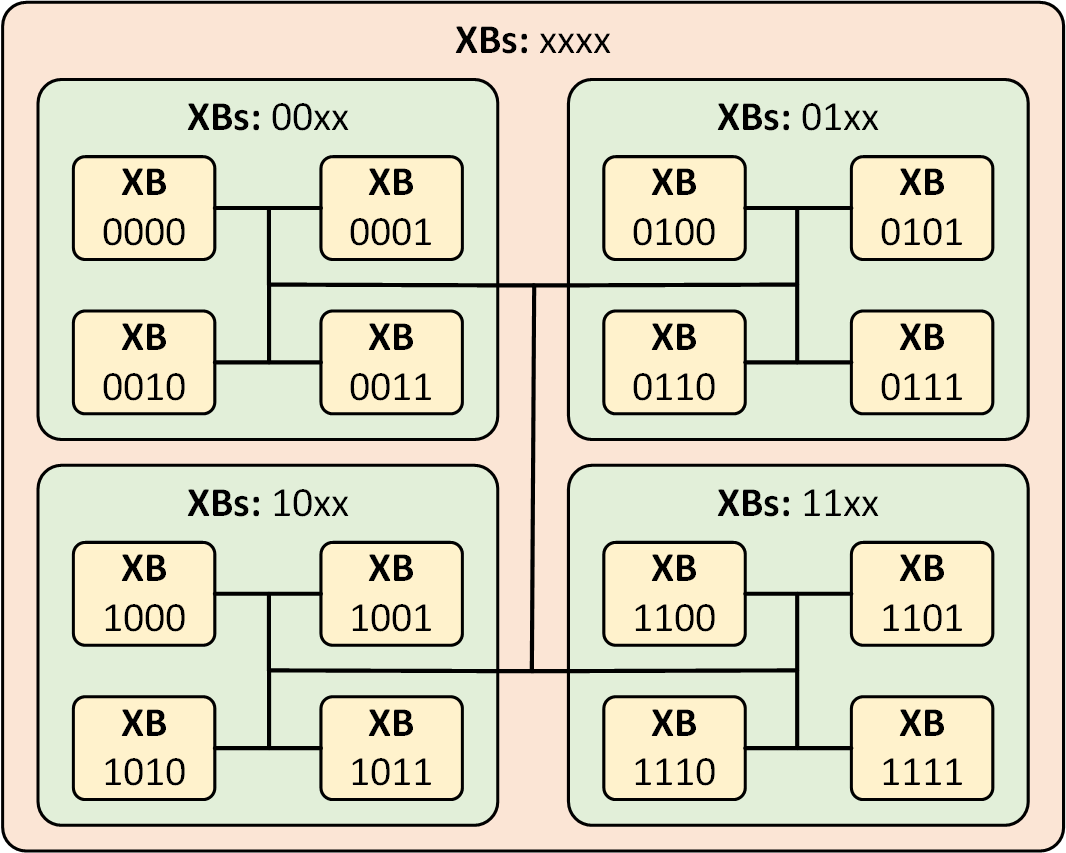}
\caption{An example hierarchical H-tree consisting of 16 crossbar arrays (XBs) numbered from $0000$ to $1111$. Each group of crossbars is characterized by crossbar indices with a shared prefix and differing suffices (e.g., group $10xx$ corresponds to crossbars $1000, 1001, 1010$ and $1011$).}
\label{fig:h-tree}
\end{figure}

We propose the recursive H-tree structure seen in Figure~\ref{fig:h-tree} (for an example of 16 crossbar arrays) that enables distributed inter-crossbar communication since each group of the hierarchy is capable of either intra-group communication (in parallel to the other groups at the same level) or inter-group communication (to other groups at the same level or higher). We have chosen an H-tree structure as opposed to other communication frameworks due to the simplicity of its implementation and its generalizability to a wide range of applications since it enables inter-crossbar communication across various levels of parallelism and inter-crossbar distance. The numbering of the crossbars in the H-tree (which corresponds to the numbering for the crossbar mask operation) is constructed recursively where each group includes all of the crossbars that share a certain prefix (e.g., group $10xx$ includes $1000, 1001, 1010$ and $1011$). Overall, the example structure in Figure~\ref{fig:h-tree} supports the following types of data movement:
\begin{itemize}
    \item Data movement within each crossbar (using Section~\ref{sec:micro:vlogic}) and in parallel across up to all crossbars.
    \item Data movement within each group of 4 crossbars, in parallel across up to all other gropus. For example, XB $0001$ transfers data to XB $0010$ at the same time that XB $0101$ transfers data to XB $0110$ and so forth. We may formalize this example as crossbars $xx01$ transferring data to crossbars $xx10$ for all $xx$. 
    \item Data movement within the group of 16 crossbars, in parallel across other groups of 16 crossbars.
\end{itemize}

We formalize the distributed communication pattern for a set of crossbar pairs as follows. Consider the set of crossbars that constitute the input crossbar of each pair, $XBs = \{XB_{start}, XB_{start}+XB_{step}, \hdots, XB_{stop}\}$ where $XB_{step}$ is a power of 4 and $XB_{stop} - XB_{start}$ is a multiple of $XB_{step}$. Let $XB_{dist}$ be the distance between the input and output crossbar of each pair -- we restrict this distance to be uniform across all pairs. Therefore, we find that each crossbar $XB$ in $XBs$ will transfer data to crossbar $XB + XB_{dist}$. For example, the transfer from the previous paragraph (crossbars $xx01$ transferring data to crossbars $xx10$ for all $xx$) will be formatted as $XB_{start} = (0001)_2, XB_{step} = (0100)_2, XB_{end} = (1101)_2$ and $XB_{dist} = (0010)_2 - (0001)_2$.

We utilize the above formalization to represent move operations in the proposed microarchitecture. To perform a distributed inter-crossbar move operation, the crossbar mask is first set to match $XB_{start}, XB_{step}, XB_{end}$ via a mask operation, and then a move operation is issued with $XB_{dist}$, the source/destination rows, and the intra-partition indices.\footnote{We store $XB_{dest} = XB_{start} + XB_{dist} \geq 0$ to avoid negative $XB_{dist}$.} The input crossbars (identified as crossbars activated by the mask) essentially perform an $N$-bit read operation that outputs the result on the H-tree bus, the output crossbars perform a write operation, and interconnect switches control the connection between the groups according to $XB_{step}$.

\begin{figure*}[!t]
\centering 
\includegraphics[width=\linewidth, trim={0cm, 0.3cm, 0cm, 0cm}]{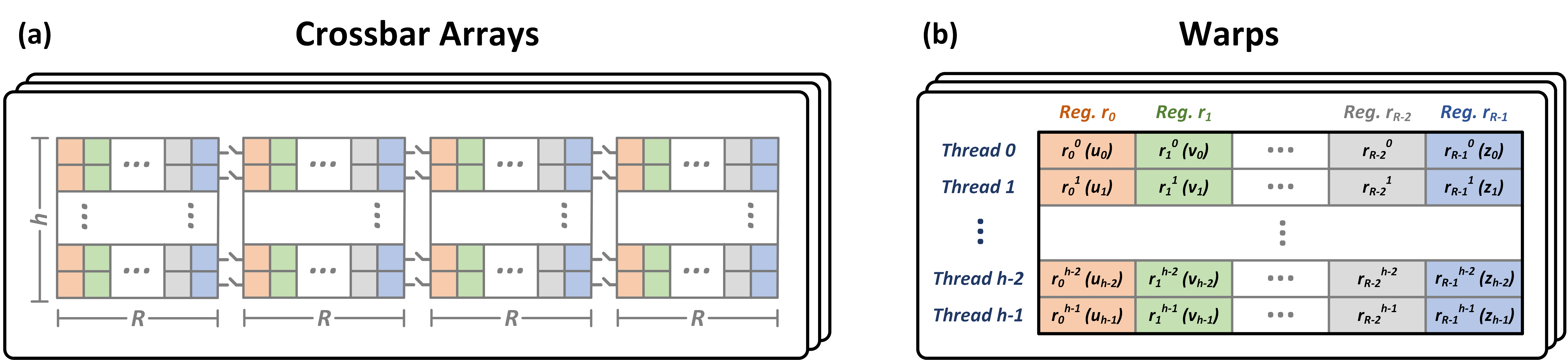}
\caption{The proposed ISA expresses (a) crossbar arrays as (b) \emph{warps} of \emph{threads} that may operate concurrently, where each register index may represent an $h$-dimensional vector (e.g., register $r_{R-1}$ represents the vector $\v{z}$).}
\label{fig:warps}
\end{figure*}

\begin{figure*}[!t]
\centering 
\includegraphics[width=\linewidth, trim={0cm, 0.3cm, 0cm, 0cm}]{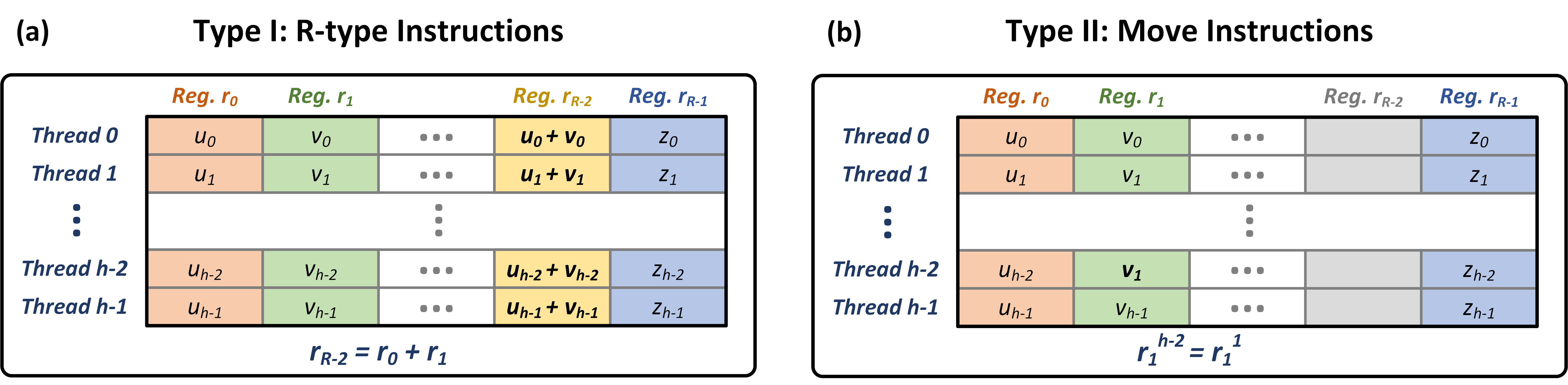}
\caption{The proposed ISA enables (a) R-type instructions (thread-parallel register operations) and (b) move instructions (inter-thread data transfer) either within warps or according to the pattern specified in Section~\ref{sec:micro:move} Note: $r_i$ refers to the register at index $i$, and $u_i,v_i,z_i$ refer to values.}
\label{fig:isa}
\end{figure*}

% ---- Instruction-Set Architecture ---- %

\section{Instruction-Set Architecture}
\label{sec:isa}

We propose a general-purpose PIM instruction set architecture (ISA) that abstracts the implementation details of memristive digital PIM. This ISA represents the interface between the proposed PIM library and the host driver, and it enables the library to generalize to routines such as logarithmic reduction while supporting all PIM architectures with an appropriate host driver. The proposed model is based on an abstraction of crossbar arrays as \emph{warps of threads}, where each thread is a single row that contains $R$ $N$-bit \emph{registers},\footnote{Where $R$ is chosen at compile-time according to $w \geq R \cdot N$, and the ISA and microarchitecture share the same word size ($N$).} as shown in Figure~\ref{fig:warps}. Warps are capable of performing R-type (register) macro-instructions (referred to as \say{instructions} for simplicity) in parallel across (up to) all threads, or inter-thread move instructions. Furthermore, inter-warp communication is also possible through \emph{move} operations between warps.

The proposed ISA differs from compute-oriented architectures (e.g., CUDA) in that the computation is performed \emph{within} the memory -- \emph{the registers of the threads are also the memory itself}. For example, consider the task of performing arithmetic on vectors stored in the memory. CUDA would \emph{allocate} threads (in dedicated CUDA cores) that copy the data from the memory to registers, perform the arithmetic on the registers, and then write the results back to the memory. Conversely, the proposed ISA would \emph{activate} the existing threads representing the memory where the vectors are stored, and those threads would operate directly on the vectors through their registers. While this may dwarf data transfer, it also imposes significant limitations on the vector alignment in the memory.

Figure~\ref{fig:isa} illustrates the two instruction types of the proposed ISA: register (R-type) and move. Register-type instructions represent arithmetic operations on the registers in parallel across (up to) all threads simultaneously, where the activated threads follow the same flexible range-based pattern defined in Section~\ref{sec:micro}. We support the AritPIM~\cite{AritPIM} suite of arithmetic functions (addition, subtraction, multiplication, and division) on both fixed-point and floating-point numbers. We further extend the ISA to include comparison operations, bitwise operations, and miscellaneous routines (e.g., absolute value, multiplexing); the full list of supported operations is available in Table~\ref{tab:isa}. Further, the ISA also supports warp-parallel thread-serial move instructions that enable communication between different threads in the same warp and communication between threads in different warps according to the inter-array move format from Section~\ref{sec:micro:move}. These instructions move a register value from one thread to a different thread when the registers are aligned, according to warp pairs that satisfy the requirements of Section~\ref{sec:micro:move}. Lastly, the ISA supports standard read and write instructions to the memory. Read instructions are targeted at a single register in a single thread of a single warp, while write instructions also target a single register yet may be repeated across several threads or warps following a range-based pattern (typically used for constants). 

\begin{table}[t]
\centering
\caption{Supported R-type operations in the proposed ISA.}
\begin{tabular}{|lcc|}
\hline
\textbf{Operation}                  & \textbf{Integer Support} & \textbf{Float Support} \\ \hline
\textbf{Arithmetic}                 &                                 &                                   \\
Addition                            & \checkmark           & \checkmark             \\
Subtraction                         & \checkmark           & \checkmark             \\
Multiplication                      & \checkmark           & \checkmark             \\
Division                            & \checkmark           & \checkmark             \\
Modulo                              & \checkmark           &                                   \\
Negation                            & \checkmark           & \checkmark             \\ \hline
\textbf{Comparison}                 &                                 &                                   \\ 
Less than (or equal to)             & \checkmark           & \checkmark             \\ 
Greater than (or equal to)          & \checkmark           & \checkmark             \\ 
Equal                               & \checkmark           & \checkmark             \\ \hline
\textbf{Bitwise}                    &                                 &                                   \\ 
Bitwise Not                         & \checkmark           & \checkmark             \\ 
Bitwise And                         & \checkmark           & \checkmark             \\ 
Bitwise Or                          & \checkmark           & \checkmark             \\ 
Bitwise Xor                         & \checkmark           & \checkmark             \\ \hline
\textbf{Miscellaneous}              &                                 &                                   \\ 
Sign                                & \checkmark           & \checkmark             \\ 
Zero                                & \checkmark           & \checkmark             \\ 
Abs                                 & \checkmark           & \checkmark             \\ 
Mux                                 & \checkmark           & \checkmark             \\ \hline
\end{tabular}
\label{tab:isa}
\vspace{-10pt}
\end{table}

% Overall, the ISA is accessed through the development library of Section~\ref{sec:libraries}. Future works may extend PyPIM to additional PIM architectures by implementing alternative host drivers that convert these ISA instructions to architecture-specific micro-operations.

% ---- PIM Library and Host Driver ---- %
\section{PIM Library and Host Driver}
\label{sec:libraries}

We propose development and driver libraries that enable PIM programming with significant ease based on the proposed ISA and microarchitecture. The development library is a Python library that provides familiar tensor bindings to the proposed ISA and is responsible for higher-level algorithms such as dynamic memory management, whereas the driver efficiently translates the ISA macro-instructions from the development library into micro-operations.

% ---- Development Library ---- %
\subsection{Development Library}
\label{sec:libraries:development}

The proposed development library enables the seamless integration of PIM into traditional tensor-based Python programs using libraries such as NumPy~\cite{NumPy} and PyTorch~\cite{PyTorch}. This both provides a familiar Python programming interface for PIM, and also enables PIM to be easily integrated within larger applications (e.g., hybrid CPU-PIM or CPU-GPU-PIM development). The library comprises the following components:
\begin{itemize}
    \item \emph{Python Bindings:} The library includes Python tensor-based bindings that enable the development of \emph{parallel} digital PIM applications using the flexibility and simplicity of Python. The example program from Figure~\ref{fig:overview} demonstrates the support for initialization of PIM vectors (e.g., \verb|pim.zeros(1024)|), direct read/write access (e.g., \verb|x[4] = 8.0|), the passing of PIM vectors as arguments (e.g., \verb|myFunc|), and parallel arithmetic (e.g., \verb|x * y|). 
    \item \emph{Dynamic Memory Management:} One of the largest challenges with digital PIM is the need for the memory to be properly aligned to enable parallelism; for example, the vectors $x$ and $y$ must be stored in the same warp for $x * y$ to be computed. The development library addresses this challenge through the combination of (1) a fall-back routine that copies vectors if they are not in the same warp (to avoid throwing an error), and (2) the careful management of PIM vectors that attempts to allocate them consecutively. Specifically, PIM vectors are allocated at a specific register index across the rows of potentially several warps, and the \verb|malloc| routine in the development library aims to map consecutive requests to the same warp ranges. 
    We support the option of providing a \emph{reference tensor} when allocating a tensor, whereby the memory allocation algorithm will first attempt to allocate memory aligned with the reference tensor. 
    \item \emph{Views and Data Movement:} We implement the slicing operation (i.e., \verb|[a:b:c]| in Python represents a slice of all numbers from inclusive $a$ to exclusive $b$ in steps of $c$) using a concept of \say{tensor views} which represent the same underlying memory. For example, when performing \verb|y = x[::2]| (select all even elements) on tensor \verb|x| then python object \verb|y| is returned which represents the same underlying memory as \verb|x| yet any operation on \verb|y| will automatically invoke a row mask corresponding to all even rows (e.g., accessing \verb|y[4]| leads to a memory read for the underlying memory of \verb|x[8]|. We generalize the inter-warp data movement supported by the ISA using these views as inter-view data-transfer is automatically converted to the underlying move operations. For example, performing \verb|x[::2] + x[1::2]| (sum of even and odd elements for a resulting tensor half the size) leads to the development library automatically identifying the move operations required to align the values before the addition (e.g., move operations will copy the contents of \verb|x[1::2]| to a new tensor aligned with \verb|x[::2]| and only then perform addition). This both enables data sharing with minimal effort, and also may serve as a powerful abstraction of inter-warp communication.
\end{itemize}

% ---- Host Driver ---- %
\subsection{Host Driver}
\label{sec:libraries:driver}

The host driver is responsible for the translation of the abstract macro-instructions (e.g., register add) into the micro-operations that adhere to the proposed microarchitecture (e.g., logic NOR). While previous works implement this step in a dedicated on-chip controller~\cite{SIMDRAM, RACER, PIMDB}, we propose an efficient host program that is not a bottleneck to PIM performance (see Section~\ref{sec:evaluation}). The efficiency is due to the combination of efficient low-level implementations of the AritPIM~\cite{AritPIM} suite of algorithms,\footnote{The algorithms were modified slightly to accommodate additional cases (e.g., signed integer division rather than only unsigned) according to the recommended extensions discussed in the AritPIM~\cite{AritPIM} library. Further, the integer multiplication algorithm was truncated to output 32 bits instead of 64.} and the compact representation for partition operations (Section~\ref{sec:micro}) that maintains moderate data transfer between the host driver and the on-chip controller. While a dedicated on-chip hardware controller would further reduce data transfer to only the encoding of the macro-instruction rather than all of the micro-operations, this would require custom controller circuitry and serve as barrier to future changes to the controller (in contrast to a software driver that may be readily updated). The development library and ISA proposed in PyPIM may also be directly applied to different digital PIM architectures by replacing the host driver, \emph{thereby enabling the same high-level PIM application to be applicable to a wide variety of digital PIM architectures}. Moreover, manufacturers of digital PIM architectures may simply adapt the PyPIM host driver according to their technologies.

% ---- End-to-End Integration ---- %
\subsection{End-to-End Integration}
\label{sec:libraries:integration}

\begin{figure}
\begin{lstlisting}
import pypim as pim
def myFunc(a: pim.Tensor, b: pim.Tensor):
    # Parallel multiplication and addition
    return a * b + a

# Tensor initialization
x = pim.zeros(2 ** 20, dtype=pim.float32)
y = pim.zeros(2 ** 20, dtype=pim.float32)
x[4], y[4] = 8.0, 0.5
x[5], y[5] = 20.0, 1.0
x[8], y[8] = 10.0, 1.0

# Custom function call
z = myFunc(x, y)
# Logarithmic-time reduction of even indices
print(z[::2].sum())  # 32.0 = 8 * 1.5 + 10 * 2
\end{lstlisting}
\caption{An example application for the end-to-end integration of PyPIM.}
\label{fig:prog}
\vspace{-10pt}
\end{figure}

We analyze the end-to-end integration of PyPIM through the example program in Figure~\ref{fig:prog}. The program is developed in Python and utilizes the \verb|pim.Tensor| class as a standalone replacement for the familiar \verb|numpy.array| class, with the additional functionality of element-wise operations that are automatically translated into digital PIM micro-operations.
\begin{itemize}
    \item \verb|pim.zeros(2 ** 20, dtype=pim.float32)|: The development library allocates 1M-element floating-point vectors $x$ and $y$ and initializes them to zeros. Specifically, the \verb|malloc| routine is executed and also a write macro-instruction is transmitted to the host driver (translated into the mask and write micro-operations).
    \item \verb|x[4] = 8.0; y[4] = 1.0|: These operations are translated by the programming interface into write macro-instructions that are then translated by the host driver into mask and write micro-operations.
    \item \verb|myFunc(x, y)|: The function call passes the tensors by reference, identical to the behavior of \verb|numpy.array|. Within the function, Python first performs \verb|a * b| by having the development library first allocate memory for the output tensor and then call the host driver to perform the multiplication instruction on the given memory addresses. The address of that intermediate is then used to perform addition with \verb|a|, leading to the desired result.
    \verb|z[::2]|: The function call receives a given tensor object and returns a view of the tensor that represents all even values. While the returned tensor view possesses the same data pointer for the memory, the elements in the tensor are accessed with the view's strides in order to replicate the behavior of a cloned tensor of even values (i.e., accessing the fourth element in the tensor will lead to the eighth element of the memory address being retrieved).
    \item \verb|z[::2].sum()|: The development library receives the tensor \emph{view} and then performs a logarithmic reduction operation on the vector to compute the overall sum. For example, in the case of \verb|z[::2].sum()|, the first step is to compute the summation of sub-tensors \verb|z[::4].sum()| and \verb|z[2::4].sum()|. This summation between non-aligned vectors is first automatically mapped to intra- and inter-warp move operations by the development library and is then followed by traditional summation on aligned tensors. This process repeats recursively until the last element is retrieved as the summation of all elements.
\end{itemize}

% ---- Evaluation ---- %
\section{Evaluation}
\label{sec:evaluation}

\begin{table}[t]
    \centering
    \caption{The evaluation parameters for PyPIM.}
    {
    \renewcommand{\arraystretch}{1.05}
    \begin{tabular}{|c|l|}
        \hline
        \textbf{System} & \textbf{Parameters} \\
        \hline
        \hline
        \multirow{4}{*}{Host} & \emph{CPU:} 2x AMD EPYC 7513 (32 core) \\
        & \emph{GPU:} NVIDIA A6000\\
        & \emph{Memory Size:} 16x 64GB \\
        & \emph{Memory BW (per chip):} 25.6 GB/s \\
        \hline
        \multirow{4}{*}{Simulated PIM~\cite{AritPIM}} & \emph{Memory Size:} 8GB \\
        & \emph{Crossbars:} $1024 \times 1024$ (32 partitions)\\
        & \emph{Word Size ($N$)}: 32\\
        % & \emph{Partition Size}: 32 bits\\
        & \emph{Clock Frequency:} 300 MHz \\
        \hline
    \end{tabular}
    }
    \label{tab:params}
\end{table}

We evaluate PyPIM to demonstrate the correctness and performance of the proposed mechanisms as compared to theoretical PIM results. To that end, we first develop a bit-accurate GPU-accelerated digital PIM simulator that is a drop-in replacement for a digital PIM chip as it performs micro-operations on an internal simulated memory in the same manner as performed by a memristive PIM chip. This enables correctness verification by (1) loading the memory with sample data, (2) performing the PyPIM algorithms with the micro-operations generated by the host driver being forwarded to the simulator, and (3) reading the results and comparing with expected values (e.g., computed using standard CPU libraries). Further, the simulator keeps track of basic profiling metrics (e.g., the number of micro-operations performed from each micro-operation type) that are then used to compare the performance of step (2) with the expected theoretical performance. Table~\ref{tab:params} lists the parameters used in the evaluation, including both the host system on which the libraries are executed and the parameters for the digital PIM architecture.

% ---- Correctness ---- %
\subsection{Correctness}
\label{sec:evaluation:correctness}

We evaluate the correctness of the proposed end-to-end integration from the high-level Python code to the generated sequence of micro-operations through a bit-level PIM simulator. The simulator is a drop-in replacement for a digital PIM chip, thereby verifying that the proposed libraries lead to the intended results when paired with a digital PIM architecture. We follow the previously-established standard of cycle-accurate simulation~\cite{MatPIM, HashPIM, FourierPIM, Ameer, AritPIM}, which requires that (1) the only interaction between the simulator and the library is through the interface for micro-operations, (2) the simulator models the operations cycle-by-cycle and executes them the same as a digital PIM chip, and (3) the results are compared to the intended output as generated by a trusted CPU-only program. Since we are simulating an entire PIM memory (in contrast to previous works that typically only simulated single rows or crossbars), we also accelerate the simulation time by using a CUDA-enabled GPU and exploiting two optimizations:
\begin{itemize}
    \item \emph{Memory:} The simulator stores the state of the digital PIM chip in the GPU memory -- the logical states of all rows of all crossbars. Whereas previous works stored each row as a vector of Booleans (effectively stored as bytes)~\cite{AritPIM, MatPIM, HashPIM, FourierPIM}, we store rows in a condensed 32-bit format that is defined according to the strided data format.
    \item \emph{Logic:} We perform the logic operations efficiently by exploiting CUDA bitwise arithmetic operations to perform semi-parallel and parallel operations (rather than iterating over the partitions explicitly). Furthermore, the micro-operations are performed in batches to improve the memory locality of each crossbar in the simulator. 
\end{itemize}

We evaluate PyPIM on the following set of benchmarks that test the different tensor operations supported by the development library (available in the code repository~\cite{repo}) and their extension to inter-crossbar algorithms:
\begin{itemize}
    \item \emph{Fundamental Arithmetic:} The four fundamental arithmetic operations (addition, subtraction, multiplication, and division) performed on randomly-generated integer and floating-point tensors.
    \item \emph{Comparison Operations:} The supported comparison operations (less than, less than or equal to, greater than, greater than or equal to, equal to, and not equal to) with randomly-generated integer and floating-point tensors.
    \item \emph{CORDIC Sine/Cosine:} We utilize the Python interface of the PyPIM library to construct sine/cosine approximations using the CORDIC algorithm~\cite{CORDIC}. We evaluate these algorithms on randomly-generated floating-point tensors in the range $[-\pi/2, \pi/2]$.
    \item \emph{Reduction:} We implement logarithmic reduction~\cite{Bitlet} using the Python interface of the PyPIM library, effortlessly performing the data movements (both intra- and inter-crossbar) through the \say{Tensor views} functionality (see Section~\ref{sec:libraries:development}). We evaluate reduction with both summation and multiplication across randomly-generated integer and floating-point tensors.
    \item \emph{Sorting:} Similar to the reduction benchmark, we utilize the Python interface of PyPIM alongside the supported \say{Tensor views} to implement intra- and inter-crossbar sorting in only 18 lines of Python code. We utilize a bitonic sorting network~\cite{Batcher} that expresses sorting as a sequence of parallel compare-and-swap operations (receive two numbers $a,b$ and output $\min(a,b), \max(a,b)$) followed by data movement between elements.
\end{itemize}

We execute these benchmarks with the proposed simulator and compare the results with those generated by the CPU-based NumPy~\cite{NumPy} library (which adheres to the IEEE 754 floating-point standards). Further, we include the implementation of these algorithms in the PyPIM library package so that they may be used directly in other works (e.g., \verb|x.sort()|).

% ---- Throughput ---- %
\subsection{Throughput}
\label{sec:evaluation:throughput}

We verify that the proposed framework attains the potential throughput provided by digital PIM for in-memory logic. We measure the performance of the framework on the unit tests discussed in the previous subsection, whereby we evaluate the following two characteristics:
\begin{itemize}
    \item \emph{PIM Cycle Count:} We compare the number of PIM cycles (number of micro-operations) performed for the unit test to the theoretical lower-bound required based on previous works (e.g., AritPIM).
    \item \emph{Host Driver Runtime:} To verify that the host driver does not bottleneck the PIM performance, we evaluated the maximal supported throughput of the host driver and compared it to the expected PIM throughput. This demonstrates the claim that a hardware controller is not necessary due to the efficiency of the host driver.
\end{itemize}
Figure~\ref{fig:results} presents these results across the benchmarks from the previous subsection,\footnote{Due to limited space, the Figure only includes a few representation benchmarks (e.g., only \say{less than} from the \say{comparison} benchmarks) with the full results available in the code repository~\cite{repo}.} displaying for each (1) the PIM performance derived from the latency metrics collected by the simulator, (2) the PIM performance derived from the theoretical latency, and (3) the maximal performance that can be supported by the host driver. We find that PyPIM is on average (worst case) $5\%$ ($16\%$) away from theoretical PIM, and that the host driver is on average (worst case) $9.5\times$ ($6.8\times$) faster than PyPIM (demonstrating that it is not a bottleneck).

\begin{figure}[t]
    \centering
    \begin{tikzpicture}
    \begin{groupplot}[
        group style={
            group size=1 by 2,
            vertical sep=1.5cm,
            x descriptions at=edge bottom,
        },
        ylabel={\footnotesize Tput. (OP/sec)},
        ymode=log,
        xticklabel style={rotate=20, anchor=north east},
        enlarge x limits=0.2,
        width=\linewidth,
        height=1.375in,
    ]
    \nextgroupplot[ybar, bar width=8pt, symbolic x coords={Int Add, Int Mult, Int $<$, FP Add, FP Mult}, title={Throughput Comparison}]
    \addplot coordinates {(Int Add,208E12)(Int Mult,17.4E12)(Int $<$,197E12)(FP Add,14.7E12)(FP Mult,12.7E12)}; \label{plots:pypim}
    \addplot coordinates {(Int Add,212E12)(Int Mult,16.1E12)(Int $<$,205E12)(FP Add,14.81E12)(FP Mult,14.31E12)}; \label{plots:theoretical}
    \addplot coordinates {(Int Add,1428E12)(Int Mult,142E12)(Int $<$,1429E12)(FP Add,112E12)(FP Mult,98E12)}; \label{plots:host}

    \coordinate (top) at (rel axis cs:0,1);
    
    \nextgroupplot[ybar, bar width=8pt, symbolic x coords={CORDIC Sine, FP Sum Reduce, FP Mult Reduce, FP Sort 1k, FP Sort 64k}, ymin=1E10, ymax=1E13]
    \addplot coordinates {(CORDIC Sine,62E9)(FP Sum Reduce,875E9)(FP Mult Reduce,762E9)(FP Sort 1k,310E9)(FP Sort 64k,52.2E9)};
    \addplot coordinates {(CORDIC Sine,66E9)(FP Sum Reduce,897E9)(FP Mult Reduce,869E9)(FP Sort 1k,326E9)(FP Sort 64k,62E9)};
    \addplot coordinates {(CORDIC Sine,5.8E12)(FP Sum Reduce,6.7E12)(FP Mult Reduce,5.9E12)(FP Sort 1k,2.1E12)(FP Sort 64k,354E9)};

    \coordinate (bot) at (rel axis cs:1,0);
    
    \end{groupplot}
    
    % Single legend
    % legend
    \path (top|-current bounding box.north)--
          coordinate(legendpos)
          (bot|-current bounding box.north);
    \matrix[
        matrix of nodes,
        anchor=south,
        draw,
        inner sep=0.2em,
        draw
      ] at([yshift=1ex]legendpos)
      {
    \ref{plots:pypim}& PyPIM&[5pt]
    \ref{plots:theoretical}& Theoretical PIM&[5pt]
    \ref{plots:host}& Host Driver&[5pt]\\};
    
  \end{tikzpicture}
  \caption{Overview of the results across the different benchmarks for the PyPIM library and the theoretical PIM performance (according to Table~\ref{tab:params}), as well as the theoretical maximal throughput supported by the host driver (demonstrating that the host driver is not a bottleneck).}
  \label{fig:results}
\end{figure}

% ---- Additional Considerations ---- %
\section{Additional Considerations}
\label{sec:misc}

We present in this section several miscellaneous considerations and limitations of PyPIM.

\begin{enumerate}

\item \textbf{Virtual Memory:} While PyPIM does not explicitly support memory virtualization, future work may add this capability by either: (1) providing software virtualization through the PyPIM libraries, or (2) utilizing the same hardware mechanisms present in traditional memory virtualization after the controller -- where each page corresponds to a single warp (e.g., 128KB pages). 

\item \textbf{Reliability and Endurance:} There is an ongoing effort to address the reliability and endurance of memristors for digital PIM. PyPIM does not exacerbate the problem with either reliability or endurance, and such reliability techniques can also be integrated into the host driver. Regardless, the same concepts proposed in PyPIM may be applied to other architectures with improved reliability or endurance such as DRAM PIM~\cite{Ambit, SIMDRAM, ComputeDRAM}.

\item \textbf{Parameter Exposure:} While the proposed ISA abstracts away the low-level details of the PIM architectures, the crossbar array dimensions dictate the number of registers per thread and the number of threads per warp. Further, PyPIM requires that the internal mechanisms and parameters of the digital PIM architecture (e.g., the supported logic gates) be exposed to the host driver. If the manufacturer does not intend to reveal this information, then the manufacturer may provide an obfuscated~\cite{Obfuscation} driver that encodes this information.

\item \textbf{Technology Simulations:} PyPIM covers the integration of PIM from high-level Python instructions to the low-level micro-operations that are issued to the memory arrays. Future work may further continue through the integration with lower-level circuit simulations of the underlying memory technology. Such simulations may be integrated with PyPIM by (1) passing the micro-operations generated by the host driver to the simulation rather than the current GPU-based logical PIM simulator, and (2) integrating the circuitry detailed in Section~\ref{sec:micro} as part of the simulations.

\end{enumerate}

% ---- Conclusion ---- %
\section{Conclusion}
\label{sec:conclusion}

As digital PIM architectures continue to emerge, there has been significant work on the algorithmic potential of the underlying parallel computing model -- from high-throughput arithmetic to large-scale applications. This paper provides the first end-to-end architectural integration of digital PIM from the low-level micro-operation format to a familiar tensor-based Python programming interface. This is accomplished by first proposing a microarchitecture for partition-enabled digital memristive PIM and an abstract instruction set architecture (ISA), and then developing libraries that enable parallel vectored PIM operations in Python programs with significant ease. We further propose a GPU-accelerated simulator that interfaces with the driver to verify correctness and enable the efficient testing of PIM applications. Overall, PyPIM increases the accessibility of PIM to the wider community through a familiar programming interface with a powerful abstraction.

\section{Acknowledgments}
This work was supported by the European Research Council through the European Union's Horizon 2020 Research and Innovation Programme under Grants 757259 and 101069336. This research received partial support from the Cisco University Research Program Fund.

\section*{Artifact Appendix}

%%%%%%%%%%%%%%%%%%%%%%%%%%%%%%%%%%%%%%%%%%%%%%%%%%%%%%%%%%%%%%%%%%%%%
\subsection{Abstract}

The artifact repository includes the framework proposed in the paper and the additional tests and benchmarks that consist of the evaluation of the framework. The framework enables high-level programming of PIM applications with significant ease as it follows the same abstraction as existing tensor-based Python libraries (e.g., NumPy, PyTorch, TensorFlow) to provide the user with a familiar interface. We utilize a CUDA-accelerated cycle-accurate simulator of the PIM framework in lieu of a physical PIM chip (requiring an NVIDIA GPU in order to test the framework) to enable the library to be tested and debugged today. The goal of this artifact repository is both to facilitate the installation and usage of the framework in order to support its integration in future works, and to present the tests and benchmarks required to replicate the key results from the work. Expected output for all the tests and benchmarks is also provided in the repository.

\subsection{Artifact check-list (meta-information)}

{\small
\begin{itemize}
  \item {\bf Compilation: }CUDA Toolkit and C++ compiler.
  \item {\bf Run-time environment: }CUDA and Python.
  \item {\bf Hardware: }NVIDIA GPU with at least 9GB DRAM.
  \item {\bf Output: }The provided tests verify correctness and performance, with expected output also provided.
  \item {\bf How much disk space required (approximately)?: }$<30GB$.
  \item {\bf How much time is needed to prepare workflow (approximately)?: }30 minutes.
  \item {\bf How much time is needed to complete experiments (approximately)?: }Approximately 4 hours.
  \item {\bf Publicly available?: }Yes.
  \item {\bf Archived: }\url{https://doi.org/10.5281/zenodo.13733284}.
\end{itemize}
}

%%%%%%%%%%%%%%%%%%%%%%%%%%%%%%%%%%%%%%%%%%%%%%%%%%%%%%%%%%%%%%%%%%%%%
\subsection{Description}

\subsubsection{How to access}

The source code can be cloned from the GitHub repository~\cite{repo}.

\subsubsection{Hardware dependencies} CUDA-capable NVIDIA GPU with at least 8GB DRAM. Tested on NVIDIA A6000.

\subsubsection{Software dependencies}
\begin{itemize}
    \item CUDA Toolkit (at least 12.0)
    \item Compiler for C++ 17 (compatible with CUDA version).
    \item Python and pip installation (tested with 3.10)
\end{itemize}
Alternatively, we recommend using the PyTorch docker container~\cite{PyTorchDocker} as it contains all of the required dependencies.

\subsubsection{Data sets} None are required as performance and correctness are measured on randomly-generated data created as part of the tests.

%%%%%%%%%%%%%%%%%%%%%%%%%%%%%%%%%%%%%%%%%%%%%%%%%%%%%%%%%%%%%%%%%%%%%
\subsection{Installation}

Build and install the library by running the following from the root directory of the repository:
\begin{lstlisting}[language=bash]
$ pip install -e .
\end{lstlisting}

Test if the installation succeeded by attempting to import the Python library and verifying that there are no errors:
\begin{lstlisting}[language=bash]
$ python
 >>> import pypim as pim
 >>>
\end{lstlisting}

%%%%%%%%%%%%%%%%%%%%%%%%%%%%%%%%%%%%%%%%%%%%%%%%%%%%%%%%%%%%%%%%%%%%%
\subsection{Experiment workflow}

The following three commands replicate the correctness and performance measurements of PyPIM (see Section~\ref{sec:evaluation} and Figure~\ref{fig:results} for further details), with their expected output available in the \emph{results} folder of the repository:
\begin{itemize}
    \item \emph{Unit tests:} \verb|python tests/unit.py|
    \item \emph{Reduction tests:} \verb|python tests/reduction.py|
    \item \emph{Sort tests:} \verb|python tests/sort.py|
\end{itemize}

To derive the maximal supported PyPIM throughput from the latency reported by the simulator we use the following equation:
\begin{equation}
\begin{split}
& Throughput [ops/sec] = \\
& Throughput[ops/cycle] \cdot Frequency[cycles/sec] = \\
& \frac{Parallelism[ops]}{Latency[cycles]} \cdot Frequency[cycles/sec]
\end{split}
\end{equation}
where $Parallelism[ops]$ is the number of rows of the crossbar memory (e.g., 64M according to Table~\ref{tab:params}). 

The performance of the host driver itself is evaluated by simulating the conditions of an ideal digital PIM chip and focusing on what is the maximal throughput of PIM operations that the PIM chip may operate at while having the host driver generate the micro-operations (i.e., what is the maximal PIM throughput supported by the host driver if the PIM chip was not the bottleneck). To that end, we model this situation by rerouting the micro-operations to a memory buffer instead of the GPU-based simulator, thereby measuring the performance of the CPU operations of the host driver involved in the micro-operation generation. Specifically, we modify the \verb|csrc/driver/driver.cpp| file as follows:
\begin{enumerate}
\item Modify the imports as follows and have the \verb|perform| function of the simulator be overridden with a write to a memory buffer (\verb|__LINE__| is used as a sequence of consecutive numbers so that each micro-operation of an instruction is written to a different address). 
\begin{lstlisting}[language=C]
// #include <pybind11/pybind11.h>
#include <iostream>
#include <chrono>
#include "driver.h"
#include "simulator.cuh"

pim::otype OPS[100000];
#define perform(x) OPS[__LINE__] = x;
\end{lstlisting}
\item Comment out the code block starting at line \verb|PYBIND11_MODULE(driver, m)| and add the benchmark test afterwards. For example, for the benchmark of integer addition, we add:
\begin{lstlisting}[language=C]
int main()
{
    pim::otype res = 0;
    std::chrono::steady_clock::time_point start = std::chrono::steady_clock::now();
    for (long i = 0; i < 10000000l; i++)
    {
        pim::add<int>(rand() % 32, rand() % 32, rand() % 32, pim::driverCrossbarMask, pim::driverCrossbarMask);
        res += OPS[rand() % 100000];
    }
    std::chrono::steady_clock::time_point end = std::chrono::steady_clock::now();
    std::cout << "Time difference = " << std::chrono::duration_cast<std::chrono::microseconds>(end - start).count() << "[us]" << std::endl;
}
\end{lstlisting}
The other supported macro-instructions can be performed by replacing e.g. \verb|pim::add| with \verb|pim::multiply| or \verb|pim::divide|, and replacing \verb|<int>| with \verb|<float>|.

\item Compile from within the \verb|csrc/driver| directory with 
\begin{lstlisting}
g++ driver.cpp -I ../ -I ../simulator/ -O3 -o driver_benchmark
\end{lstlisting}
and run \verb|./driver_benchmark|.
\end{enumerate}

%%%%%%%%%%%%%%%%%%%%%%%%%%%%%%%%%%%%%%%%%%%%%%%%%%%%%%%%%%%%%%%%%%%%%
\subsection{Evaluation and expected results}

The expected results of the above commands are available in the \emph{results} directory of the repository (the exact output of the tests, reflecting the performance measurements from the GPU-based simulator). The provided tests only succeed if the correctness comparison of the framework with the expected ground-truth result (based on NumPy) is successful; for example, see the \emph{test\_arit} test (from the \emph{tests/unit.py} file) which verifies the correctness of all arithmetic operations on integer and floating-point numbers:

\begin{lstlisting}
@parameterized.expand([
 ('__add__', np.add, np.dtype(np.int32)),
 ('__sub__', np.subtract, np.dtype(np.int32)),
 ('__mul__', np.multiply, np.dtype(np.int32)),
 ('__truediv__', np.true_divide, np.dtype(np.int32)),
 ('__add__', np.add, np.dtype(np.float32)),
 ('__sub__', np.subtract, np.dtype(np.float32)),
 ('__mul__', np.multiply, np.dtype(np.float32)),
 ('__truediv__', np.true_divide, np.dtype(np.float32))
]) 
def test_arit(self, function, gt_func, dtype):
    print(f'Testing arithmetic with {function} on type {dtype}')
    nelem = 2 ** 16
    ninputs = 2
    
    # Allocate and initialize the tensors
    refs = [utils.rand(nelem, dtype) for _ in range(ninputs)]
    tensors = [pim.from_numpy(x) for x in refs]
    
    # Perform the arithmetic function
    with pim.Profiler():
        result = getattr(tensors[0], function)(tensors[1])
    result = pim.to_numpy(result)

    # Compare to ground-truth
    ground_truth = gt_func(refs[0], refs[1]).astype(dtype)
    np.testing.assert_array_equal(ground_truth, result)
\end{lstlisting}

%%%%%%%%%%%%%%%%%%%%%%%%%%%%%%%%%%%%%%%%%%%%%%%%%%%%%%%%%%%%%%%%%%%%%
\subsection{Experiment customization}

The experiments may be customized through modifications to the testing parameters such as \verb|nelem| in the above scripts. We further encourage the overall experimentation with the library in an interactive format as with existing tensor-based Python libraries, or using custom tests by following the same file structure as the provided tests. For example, see the following interactive use of the framework through the standard Python interpreter which tests tensor allocation, read/write operations, tensor views, summation reduction and sorting:

\begin{lstlisting}
>>> import pypim as pim
>>> x = pim.zeros(8, dtype=pim.float32)
>>> x
Tensor(shape=(8,), dtype=float32):
  [0.0, 0.0, 0.0, 0.0, 0.0, 0.0, 0.0, 0.0]
>>> 
>>> x[2] = 2.5
>>> x[3] = 1.25
>>> x[4] = 2.25
>>> x
Tensor(shape=(8,), dtype=float32):
  [0.0, 0.0, 2.5, 1.25, 2.25, 0.0, 0.0, 0.0]
>>> 
>>> x[::2]
TensorView(shape=(4,), dtype=float32, slicing=slice(0, 7, 2)):
  [0.0, 2.5, 2.25, 0.0]
>>> x[::2].sum()
4.75
>>> x[::2].sort()
TensorView(shape=(4,), dtype=float32, slicing=slice(0, 7, 2)):
  [0.0, 0.0, 2.25, 2.5]
\end{lstlisting}

%%%%%%%%%%%%%%%%%%%%%%%%%%%%%%%%%%%%%%%%%%%%%%%%%%%%%%%%%%%%%%%%%%%%%
\subsection{Notes}

The derivation of the throughput results of Figure~\ref{fig:results} follows from the latency and gate count metrics produced by the repository according to the constants from Table~\ref{tab:params}.

%%%%%%%%%%%%%%%%%%%%%%%%%%%%%%%%%%%%%%%%%%%%%%%%%%%%%%%%%%%%%%%%%%%%%
\subsection{Methodology}

Submission, reviewing and badging methodology:

\begin{itemize}
  \item \url{https://www.acm.org/publications/policies/artifact-review-and-badging-current}
  \item \url{https://cTuning.org/ae}
\end{itemize}

%%%%%%%%%%%%%%%%%%%%%%%%%%%%%%%%%%%%%%%%%%%%%%%%%%%%
% When adding this appendix to your paper, 
% please remove below part
%%%%%%%%%%%%%%%%%%%%%%%%%%%%%%%%%%%%%%%%%%%%%%%%%%%%

% ---- References ---- %
\balance
\bibliographystyle{IEEEtranS}
\bibliography{refs}

\end{document}